\documentclass[12pt]{iopart}

\expandafter\let\csname equation*\endcsname\relax
\expandafter\let\csname endequation*\endcsname\relax
\usepackage{graphicx}% Include figure files
\usepackage{iopams}
\usepackage{ytableau}
\usepackage{tikz}

\usepackage{mathtools}
\usepackage{relsize}
\DeclarePairedDelimiter\bra{\langle}{\rvert}
\DeclarePairedDelimiter\ket{\lvert}{\rangle}
\DeclarePairedDelimiterX\braket[2]{\langle}{\rangle}{#1 \delimsize\vert #2}

\usepackage{subfig}
\usepackage{multirow}
\usepackage{pdflscape}

\usepackage{cite}
\allowdisplaybreaks
\begin{document}

%\title[Sample title]{Multi-level generalization of the Tavis-Cummings model: applications to anharmonic vibrational polaritons}
\title[Multi-level Tavis-Cummings]{Generalization of the Tavis-Cummings model for multi-level anharmonic systems.}%Multi-level generalization of the Tavis-Cummings model: applications to anharmonic vibrational polaritons}
% Force line breaks with \\

\author{J. A. Campos-Gonzalez-Angulo$^1$, R. F. Ribeiro$^2$ and J. Yuen-Zhou$^1$}
 \address{$^1$ Department of Chemistry and Biochemistry. University of California San Diego. La Jolla, California 92093, USA}%Lines break automatically or can be forced with \\
\address{$^2$ Department of Chemistry. Emory University. Atlanta, Georgia 30322, USA}%
\ead{joelyuen@ucsd.edu}
% \homepage{http://yuenzhougroup.ucsd.edu}


\begin{abstract}
The interaction between anharmonic quantum emitters (e.g., molecular vibrations) and confined electromagnetic fields gives rise to quantum states with optical and chemical properties that are different from those of their precursors. The exploration of these properties has been typically constrained to the first excitation manifold, the harmonic approximation, ensembles of two-level systems [Tavis-Cummings (TC) model], or the anharmonic single-molecule case. The present work studies, for the first time, a collective ensemble of identical multi-level anharmonic emitters and their dipolar interaction with a photonic cavity mode. The permutational properties of the system allow identifying symmetry classified submanifolds in the energy spectrum. Notably, in this approach, the number of particles, typically in the order of several millions, becomes only a parameter from the operational standpoint, and the size of the dimension of the matrices to diagonalize is independent of it. The formalism capabilities are illustrated by showing the energy spectrum structure, up to the third excitation manifold, and the calculation of the photon contents as a permutationally invariant quantity. Emphasis is placed on (a) the collective (superradiant) scalings of light-matter couplings and the various submanifolds of dark (subradiant) states with no counterpart in the single-molecule case, as well as (b) the delocalized modes containing more than one excitation per molecule with no equivalent in the TC model. We expect these findings to be applicable in the study of non-linear spectroscopy and chemistry of polaritons.
\end{abstract}
%\submitto{\NJP}
%\maketitle

\section{Introduction}

Optical microcavities and similar devices enable the dipolar interaction between the electromagnetic (EM) field they confine and a suitable degree of freedom of quantum emitters\cite{Ebbesen2016,RibeiroMartinez-MartinezDuEtAl2018,FlickRiveraNarang2018,FeistGalegoGarcia-Vidal2018,HerreraOwrutsky2020}. The coupled system, whose excitations receive the name of polaritons, is a hybrid between light and matter, and its properties have motivated extensive exploration from disciplines such as quantum optics\cite{Angelakis2017}, excitonic and two-dimensional materials science\cite{AlcarazIranzoNanotDiasEtAl2018,LeeLeongKalashnikovEtAl2020,ProsciaJayakumarGeEtAl2020}, and chemistry \cite{HerreraSpano2016,HiraiHutchisonUji-i2020}. In many systems of interest, observable effects emerge only when a large number of material dipoles cooperatively interact with a single photon mode.\cite{Hopfield1958,Agranovich1960}.

Polaritons have been produced using diverse setups and materials whose description requires various theoretical frameworks. For instance, real spins, NV centers \cite{KuboOngBertetEtAl2010}, qubits \cite{AlickiHorodeckiHorodeckiEtAl2004}, SQUIDS \cite{FinkBianchettiBaurEtAl2009}, quantum dots \cite{KabussCarmeleBrandesEtAl2012,AbdussalamMachnikowski2014}, and electronic transitions in organic molecules \cite{RichterGeggTheuerholzEtAl2015} are approched as two-level systems, while low vibrational transitions in molecules can be modeled with harmonic oscillators \cite{AhnVurgaftmanDunkelbergerEtAl2018,MukamelNagata2011}. The success of these two models stems from their simplicity and the fact that conditions can be found for their realization in interesting experimental settings.%conditions for the light-matter interaction can be found for which their formulations give rise to similarly simple analytical solutions. 

The interest in strongly coupled systems in which the emitters have a multi-level anharmonic spectrum has recently increased. For instance, the role that light-matter interaction might play on chemical reactivity has been explored by studying the Rabi model with a single emitter described by a Morse oscillator \cite{HernandezHerrera2019,TrianaHernandezHerrera2020}. Additionally, signatures of the non-linearities present in states with two energy quanta have been observed experimentally \cite{XiangRibeiroDunkelbergerEtAl2018,XiangRibeiroLiEtAl2019,AutryNardinSmallwoodEtAl2020,DelPoKudischParkEtAl2020}, studied from theory \cite{SaurabhMukamel2018,RibeiroDunkelbergerXiangEtAl2018}, and motivated innovative theoretical ideas for signal enhancement \cite{RibeiroCampos-Gonzalez-AnguloGiebinkEtAl2020,GuMukamel2020}.

%On the other hand, multi-level system descriptions had been used to study electrically induced transparency\cite{FleischhauerImamogluMarangos2005}, entangled photon creation\cite{BensonSantoriPeltonEtAl2000,AkopianLindnerPoemEtAl2006,DousseSuffczynskiBeveratosEtAl2010,CarmeleKnorr2011,SchumacherFoerstnerZrennerEtAl2012,HeinSchulzeCarmeleEtAl2014}, noise-induced coherences\cite{TscherbulBrumer2014,ScullyChapinDorfmanEtAl2011}, and low threshold laser theories\cite{Haken1970,MuSavage1992}. Furthermore, they are suitable to describe vibrational polaritons with high enough excitations that the harmonic approximation, which assumes equally spaced energy levels, breaks down\cite{SaurabhMukamel2018,RibeiroDunkelbergerXiangEtAl2018,HernandezHerrera2019}. That situation is commonplace in chemical reactions, which are thermally activated processes that usually require the reactive normal mode to access excited states where the anharmonicity becomes significant\cite{Teitelbaum1993,SansonSanchezCorchado2006}.

The dynamics of the interaction between many multi-level systems and several EM modes has been discussed when non-energy-conserving terms of the Hamiltonian can be neglected, i.e., under the so-called rotating wave approximation (RWA) \cite{Skrypnyk2015,Skrypnyk2008,LeeLinksZhang2011}. From these studies, it is known that analytical solutions exist for the coupling of $r$-level systems to $r-1$ EM modes \cite{Skrypnyk2018}. In the context of the problem of $N$ molecules coupled to a single EM mode, the RWA allows separating the Hamiltonian of the system allowing numerical approaches to the computation of exact solutions. However, in this kind of systems the Hilbert space scales as the number of molecules considered; therefore, these calculations are suitable only for systems of relatively small size. Nonetheless, for the case of identical emitters, the permutational symmetry implies a high degree of degeneracy, which can be exploited to reduce the number of degrees of freedom of the Hamiltonian, thus increasing the computational capabilities of the model. Such an approach has been utilized to provide a formulation for open quantum systems that can be used to explore spectroscopic measurements and other processes involving relaxation\cite{GeggRichter2016,ShammahAhmedLambertEtAl2018}. There is, however, little to no word about the stationary features of such systems, which are discussed in depth in the present work.

Permutational symmetry has been exploited in the few body limit to study systems such as nuclear structure \cite{BohrMottelson1998}, quantum circuits \cite{BaconChuangHarrow2006}, magnons \cite{JakubczykKravetsJakubczyk2011}, ultracold atoms \cite{Harshman2016}, nuclear spins within molecules \cite{SchmiedtJensenSchlemmer2016}, and the one-dimensional Hubbard model \cite{Jakubczyk2020}. In the general case, the properties of the symmetrized quantum states have also been extensively discussed \cite{KlimovChumakov2009,Gilmore1972}. However, to the best of our knowledge, there has not been an effort to present concrete results from the application of this algebraic approach to the problem of light-matter coupling in the collective regime beyond the case of two-level systems, which is widely known \cite{ScullyZubairy1997,KlimovChumakov2009,DukalskiBlanter2013}.

In the present work, we provide a fathomable, insightful and easily implementable procedure to simplify the time-independent Schrödinger equation for $N$ emitters, each with $r$ non-degenerate and non-equispaced bound states in their energy spectrum, coupled to a harmonic field mode through a bilinear and excitation-conserving interaction. We identify a correlation between the symmetries of the system and the distribution of photonic component among the emerging manifolds. Of particular interest is the realization that collective states comprising excitations of distinct natures give rise to cooperative couplings that depart from the well-known factor of $\sqrt{N}$, which characterizes polaritons in the singly excited manifold. We also present the effects of anharmonicity, detuning, and intensity of collective coupling on the eigenenergies, and photon contents of the eigenstates.

This paper is organized as follows: in \sref{sec:formalism} we introduce the model Hamiltonian, discuss the implied approximations, and provide a layout of the algebraic structure that enables its separation. In \sref{sec:permsymm} we introduce the tools and concepts that enable the block-diagonalization, and show the form and distribution of the diagonal elements. In \sref{sec:colcoupl} we formulate the off-diagonal matrix elements that result from the block-diagonalization. In \sref{sec:algor} we summarize the simplification method, providing a general recipe to put it in practice. In \sref{sec:examp} we present examples for the separation for the lowest-lying excitation manifolds. In \sref{sec:obser} we explore the implications of varying parameters on properties of the eigenstates. Finally, we present the conclusions in \sref{sec:conclu}.

\section{\label{sec:formalism} Description of the model}

The Hamiltonian
\begin{equation}\label{eq:hamb}
\hat H_{\textrm{bare}}(N)=\hbar \omega \left(\hat a_0^\dagger \hat a_0+\frac{1}{2}\right)+\sum_{i=1}^N\sum_{v=0}^{r}\varepsilon_v \hat\sigma_{v,v}^{(i)}\enskip,
\end{equation}
describes a collection of $N$ identical emitters with $r+1$ states each (which could represent the lowest vibrational bound states of the ground electronic state of an anharmonic molecule) and an EM mode with frequency $\omega$. Here, $\hat a_0^{(\dagger)}$ is the annihilation (creation) operator of the EM mode, $\varepsilon_v$ is the energy of the $v$th  state of the emitter, and $\hat\sigma_{v,u}^{(i)}$ is the transition operator between levels $u$ and $v$ in the $i$th emitter, i.e.,
\begin{equation}\label{eq:transop}
\hat\sigma_{\alpha,\beta}^{(i)}\ket{v_0v_1\ldots v_i\ldots v_N}=\delta_{v_i \beta}\ket{v_0v_1\ldots \alpha_i\ldots v_N}.
\end{equation}
In \eref{eq:transop}, the eigenstates of $\hat H_\textrm{bare}(N)$ in \eref{eq:hamb}, henceforth primitive bare-eigenstates (PBEs), are written in a collective Fock representation where the index $v_i$ indicates the number of bosonic excitations in the $i$th emitter or the EM mode ($i=0$). Hereafter, modes with $v=0$ are considered implicitly for brevity.

The Hamiltonian describing the emitters, the EM mode, and their dipolar interactions under the RWA is
\begin{equation}\label{eq:hamt}
\hat H(N)=\hat H_{\textrm{bare}}(N)+\hat H_{\textrm{int}}(N),
\end{equation}
where
\begin{equation}\label{eq:hamint}
\hat H_{\textrm{int}}(N)=\sum_{v=0}^{r-1}\left(g_{v,v+1} \hat a_0^\dagger \hat J_{-}^{(v+1,v)} + g_{v+1,v}\hat J_{+}^{(v+1,v)} \hat a_0 \right),
\end{equation}
with collective ladder operators for the emitters,
%\numparts
\begin{equation}\label{eq:collectladder}
%\hat J_{+}^{(u,v)}=\sum_{i=1}^N \hat\sigma_{u,v}^{(i)},
\hat J_{\pm}^{\left(v+\frac{1\pm1}{2}u,v-\frac{1\mp1}{2}u\right)}=\sum_{i=1}^N \hat\sigma_{v\pm u,v}^{(i)}\qquad u>0.
\end{equation}
%and
%\begin{equation}
%\hat J_{-}^{(u,v)}=\sum_{i=1}^N \hat\sigma_{v,u}^{(i)},
%\end{equation}
%\endnumparts
Here, the coupling constant $g_{v\pm u,v}=\sqrt{\frac{\hbar \omega}{2 \epsilon_0\cal{V}}}\bra{v\pm u}e\hat x\ket{v}$ is the product of the amplitude of the single-photon electric field, confined to a mode volume $\cal{V}$, and the value of the transition dipole moment between emitter states $v\pm u$ and $v$. Notice, in \eref{eq:hamint}, that the model only accounts for coupling between states that differ by one excitation, and therefore neglects overtone transitions.

The operator $\hat n_\textrm{exc}=\hat a_0^\dagger \hat a_0 + \sum_{i=1}^N\sum_{v=0}^{r}v\hat\sigma_{v,v}^{(i)}$ acts on the PBEs according to
\begin{equation}
\hat n_\textrm{exc}\ket{v_0 v_1\ldots v_N}=\sum_{j=0}^N v_j\ket{v_0 v_1\ldots v_N},
\end{equation}
thus indicating the total number of excitations of a given state; moreover, since $\left[\hat H(N),\hat n_\textrm{exc}\right]=0$, this operator corresponds to a constant of motion. Therefore, the Hamiltonian can be recast in the form
\begin{equation}
\hat H(N)=\sum_{n_\textrm{exc}}\hat H_{n_\textrm{exc}}(N),
\end{equation}
in which each of the terms in the sum of the right-hand-side can be represented as block matrices spanned by PBEs of constant $n_\textrm{exc}$, these blocks generate the so-called \textit{excitation manifolds}. Explicitly, we have

\numparts
\begin{eqnarray}\label{eq:Hman}
\fl \hat H_0(N)=\left(\frac{\hbar\omega}{2}+N\varepsilon_0\right)\ket{0}\bra{0},\\
\fl\hat H_1(N)=\left(\frac{3}{2}\hbar\omega+N\varepsilon_0\right) \ket{1_0}\bra{1_0}+\left(\frac{\hbar\omega}{2}+\varepsilon_1+(N-1)\varepsilon_0\right)\sum_{i=1}^N \ket{1_i}\bra{1_i}\nonumber\\
+\sum_{i=1}^N\left(g_{01}\ket{1_0}\bra{1_i}+g_{10}\ket{1_i}\bra{1_0}\right),\label{eq:h1}\\
\fl\hat H_2(N)=\left(\frac{5}{2}\hbar\omega+N\varepsilon_0\right)\ket{2_0}\bra{2_0}+\left(\frac{3}{2}\hbar\omega+\varepsilon_1+(N-1)\varepsilon_0\right)\sum_{i=1}^N\ket{1_0 1_i}\bra{1_0 1_i}\nonumber\\
+\left(\frac{\hbar\omega}{2}+2\varepsilon_1+(N-2)\varepsilon_0\right)\sum_{i=1}^{N-1}\sum_{j=i+1}^N\ket{1_i 1_j}\bra{1_i 1_j}\nonumber\\
+\left(\frac{\hbar\omega}{2}+\varepsilon_2+(N-1)\varepsilon_0\right)\sum_{i=1}^N\ket{2_i}\bra{2_i}\nonumber\\
+\sum_{i=1}^N\left[\sqrt{2}g_{01}\ket{2_0}\bra{1_0 1_i}+\ket{1_0 1_i}\left(g_{12}\bra{2_i}+g_{01}\sum_{j\neq i}\bra{1_i 1_j}\right)+\textrm{H.c.}\right],\label{eq:h2}\\
\fl\hat{H}_3(N)=\left(\frac{7}{2}\hbar\omega+N\varepsilon_0\right)\ket{3_0}\bra{3_0}+\left(\frac{5}{2}\hbar\omega+\varepsilon_1+(N-1)\varepsilon_0\right)\sum_{i=1}^N\ket{2_0 1_i}\bra{2_0 1_i}\nonumber\\
+\left(\frac{3}{2}\hbar\omega+\varepsilon_2+(N-1)\varepsilon_0\right)\sum_{i=1}^N\ket{1_0 2_i}\bra{1_0 2_i}\nonumber\\
+\left(\frac{3}{2}\hbar\omega+2\varepsilon_1+(N-2)\varepsilon_0\right)\sum_{i=1}^{N-1}\sum_{j=i+1}^N\ket{1_0 1_i 1_j}\bra{1_0 1_i 1_j}\nonumber\\
+\left(\frac{\hbar\omega}{2}+\varepsilon_3+(N-1)\varepsilon_0\right)\sum_{i=1}^N\ket{3_i}\bra{3_i}\nonumber\\
+\left(\frac{\hbar\omega}{2}+\varepsilon_2+\varepsilon_1+(N-2)\varepsilon_0\right)\sum_{i=1}^N\sum_{j\neq i}\ket{2_i 1_j}\bra{2_i 1_j}\nonumber\\
+\left(\frac{\hbar\omega}{2}+3\varepsilon_1+(N-3)\varepsilon_0\right)\sum_{i=1}^{N-2}\sum_{j=i+1}^{N-1}\sum_{k=j+1}^N\ket{1_i 1_j 1_k}\bra{1_i 1_j 1_k}\nonumber\\
+\sum_{i=1}^N\left[\sqrt{3}g_{01}\ket{3_0}\bra{2_0 1_i}+\sqrt{2}\ket{2_0 1_i}\left(g_{12}\bra{1_0 2_i}+g_{01}\sum_{j\neq i}\bra{1_0 1_i 1_j}\right)+\textrm{H.c.}\right]\nonumber\\
+\sum_{i=1}^{N-1}\sum_{j=i+1}^N\left\{\ket{1_0 1_i 1_j}\left[g_{12}\left(\bra{2_i 1_j}+\bra{2_j 1_i}\right)+g_{01}\sum_{k\neq i,j}\bra{1_i 1_j 1_k}\right]+\textrm{H.c.}\right\}\nonumber\\
+\sum_{i=1}^N\left[\ket{1_0 2_i}\left(g_{23}\bra{3_i}+g_{01}\sum_{j\neq i}\bra{2_i 1_j}\right)+\textrm{H.c.}\right],\label{eq:h3}\\
\vdots\nonumber
\end{eqnarray}
\endnumparts
where $\ket{0}=\ket{0_0 0_1\ldots 0_N}$ is the global ground state and H.c. stands for Hermitian conjugate.

While  \eref{eq:Hman} represents a significant simplification that allows diagonalization of  \eref{eq:hamt}, it is important to remark that the matrices generated with this approach can quickly become intractable as the number of molecules increases. For instance, $\hat H_{n_\textrm{exc}}(N)$ is an operator in $\binom{N+n_\textrm{exc}}{n_\textrm{exc}}$ dimensions. Therefore, it is impractical to deal with the number of molecules required to achieve observable ensemble strong coupling, which is usually in the order of millions \cite{AgranovichLitinskaiaLidzey2003,PinoFeistGarcia-Vidal2015,DaskalakisMaierKena-Cohen2017}.

\section{Permutational symmetry \label{sec:permsymm}}

Further simplification of  \eref{eq:hamt} requires to identify the additional symmetries of the Hamiltonian. An inspection of  \eref{eq:Hman} reveals that PBEs with the same distribution of quanta are degenerate under the action of $\hat H_{\textrm{bare}}(N)$; therefore, it becomes convenient to introduce a label that characterizes PBEs yet avoids spurious identification due to accidental degeneracies. In principle, this characterization could be achieved with the bare energy and the collection of eigenvalues of the operators
\begin{equation}
\hat{J}_0^{(u,v)}=\frac{1}{2}\left[\hat{J}_{+}^{(u,v)},\hat{J}_{-}^{(u,v)}\right]\qquad0\leq v<u\leq r,
\end{equation}
which commute with $\hat{H}_\textrm{bare}(N)$; indeed, the relevance of these operators and those in \eref{eq:collectladder} will be revisited in \sref{sec:colcoupl}. Nonetheless, a much simpler approach is the description of the distribution of quanta itself. Let us define the \emph{spectral configuration}:
\begin{equation}\label{eq:specconf}
\tilde{\mu}=\prod_{v=0}^{r}v^{n_{\tilde{\mu}}(v)},
\end{equation}
a notation device that indicates the number of emitters $n_{\tilde\mu}(v)$ in the $v$th excited state. See \tref{tab:speconf} for usage examples.

\begin{table}
%\centering
\ytableausetup{boxsize=8pt,centertableaux}
\caption{\label{tab:speconf}Examples of spectral configurations, partitions, and Young diagrams corresponding to selected PBEs.}
\begin{indented}
\item[]\begin{tabular}{@{}lllll}
\br
%\ytableausetup{boxsize=8pt,centertableaux}
PBE&$\tilde\mu$&$\bmu$&$\bnu$&Young diagram$^{\rm a}$\\
\mr%\hline
$\ket{0}$&$0^N$&$[N]$&$[0]$&$\ydiagram{6}$\\
\bs
$\ket{1_i}$&$0^{N-1}1^1$&$[N-1,1]$&$[1]$&$\ydiagram{5,1}$\\
\bs
$\ket{1_i 1_j}$&$0^{N-2}1^2$&$[N-2,2]$&$[1,1]$&\ydiagram{4,2}\\
\bs
$\ket{2_i}$&$0^{N-1}2^1$&$[N-1,1]$&$[2]$&\ydiagram{5,1}\\
\bs
$\ket{2_i4_j5_k}$&$0^{N-3}2^1 4^1 5^1$&$[N-3,1,1,1]$&$[5,4,2]$&\ydiagram{3,1,1,1}\\
\bs
$\ket{1_i1_j2_k2_\ell2_m4_n}$&$1^2 2^3 4^1$&$[3,2,1]$&$[4,2,2,2,1,1]$&\ydiagram{3,2,1}\\
\br
%\endtab
\end{tabular}
\item[] $^{\rm a}{N=6}$
\end{indented}
\end{table}
The spectral configuration labels not only PBEs, but also any bare eigenstate (BE) resulting from linear combinations of PBEs with the same distribution of quanta.  The BEs can thus be partially characterized with three labels: the excitation manifold ($n_\textrm{exc}$), the spectral configuration ($\tilde\mu$), and a basis-dependent degeneracy index ($y$). These states fulfill
\begin{equation}
\hat{H}_\textrm{bare}(N)\ket{n_\textrm{exc},\tilde\mu;y}=\varepsilon_{n_\textrm{exc}}^{(\tilde\mu)}\ket{n_\textrm{exc},\tilde\mu;y},
\end{equation}
where
\begin{equation}\label{eq:barenerg}
\varepsilon_{n_\textrm{exc}}^{(\tilde\mu)}=\hbar\omega\left(v_0^{(n_\textrm{exc},\tilde\mu)}+\tfrac{1}{2}\right)+\sum_{v=0}^{n_\textrm{exc}}n_{\tilde\mu}(v)\varepsilon_v
\end{equation}
is the characteristic bare energy for states with those labels, and $v_0^{(n_\textrm{exc},\tilde\mu)}=n_\textrm{exc}-\sum_{v=1}^{n_\textrm{exc}}v n_{\tilde\mu}(v)$ is the corresponding number of quanta in the EM mode.

Since $\sum_{v=0}^{r}n_{\tilde\mu}(v)=N$ for a given $\tilde\mu$, the spectral configuration corresponds to a partition of the total number of emitters. For reasons that will become apparent, it is convenient to write these partitions in their regular (non-increasing) form, i.e.,
\begin{equation}
{\bmu}=[\mu_1,\mu_2,\ldots,\mu_{r+1}]
\end{equation}
where $\mu_i=n_{\tilde\mu}(v_i)$ under the constraint that $\mu_i \leq \mu_{i'}$ for $i>i'$ \cite{Andrews1998}. In what follows, all the empty elements, $\mu_i=0$, will be omitted for brevity, as illustrated in \tref{tab:speconf}. Furthermore, the spectral configurations can be identified as partitions of the number of excitations distributed among the emitters, i.e. for every $\tilde\mu$ there exists a regular partition $\bnu(\tilde\mu)$ such that
\begin{equation}
\bnu(\tilde\mu)=\{v^{(i)}:1\leq v\leq r,0\leq i\leq n_{\tilde\mu}(v)\},
\end{equation}
as illustrated in \tref{tab:speconf}. Because of the latter, the number of possible spectral configurations within a given manifold is
\begin{equation}\label{eq:specconfpart}
\lvert\tilde\mu_{n_\textrm{exc}}\rvert=\sum_{n=0}^{n_\textrm{exc}}p(n),
\end{equation}
where $p(n)$ is the number of partitions of the integer $n$. These numbers can be extracted from the expansion \cite{AbramowitzStegun1970}
\begin{equation}\label{eq:partseries}
\prod_{k=1}^m\left(1-x^k\right)^{-1}=\sum_{n=0}^m{p(n)x^n}+\Or\left(x^{m+1}\right).
\end{equation}

%\begin{table}
%\centering
%\caption{Spectral configurations, EM mode excitations and degeneracies for the first four excitation manifolds\label{tab:specdeg}}
%\begin{indented}
%\item[]\begin{tabular}{*{10}{c}}
%$\tilde\mu$&$0^N$&$0^{N-1}1^1$&$0^{N-2}1^2$&$0^{N-1}2^1$&$0^{N-3}1^3$&$0^{N-2}1^1 2^1$&$0^{N-1}3^1$&\multirow{2}{*}{$\lvert\tilde\mu_{n_\textrm{exc}}\rvert$}&\multirow{2}{*}{$\lvert{\lambda}_{n_\textrm{exc}}\rvert$}\\
%\cline{1-8}
%\rule{0pt}{12pt}{$n_\textrm{exc}$}&\multicolumn{7}{c}{Excitations in the EM mode}&\\
%\hline
%0&0&&&&&&&1&1\\
%1&1&0&&&&&&2&2\\
%2&2&1&0&0&&&&4&3\\
%3&3&2&1&1&0&0&0&7&5\\
%\cline{1-8}
%\rule{0pt}{12pt}$\dim(M^{\bmu})$&1&$N$&$\binom{N}{2}$&$N$&$\binom{N}{3}$&$2\binom{N}{2}$&$N$\\
%\rule{0pt}{12pt}$\dim(S^{\blambda})$\footnote{$\blambda=\bmu$ when applicable.}&1&$N-1$&$\frac{N}{2}(N-3)$&&$\frac{N-5}{N-2}\binom{N}{3}$&$\binom{N-1}{2}$&
%\end{tabular}
%\end{indented}
%\end{table}
\begin{table}
\footnotesize
%\centering
\caption{\label{tab:specdeg}Spectral configurations, EM mode excitations and degeneracies for the first five excitation manifolds.}
%\lineup
%\begin{indented}
%\item[]
%\begin{tabular*}{\textwidth}{@{}*{8}{l}{@{\extracolsep{0pt plus 12pt}}l}}
\begin{tabular*}{\textwidth}{@{}l*{5}{@{\extracolsep{0pt plus 12pt}}r}*{2}{@{\extracolsep{0pt plus 12pt}}l}}
\br
%\multirow{2}{*}{$\tilde\mu$}&$n_\textrm{exc}=0$&$n_\textrm{exc}=1$&$n_\textrm{exc}=2$&$n_\textrm{exc}=3$&$n_\textrm{exc}=4$&\multirow{2}{*}{$\dim(M^{\bmu})$}&\multirow{2}{*}{$\dim(S^{\blambda})$\footnote{$\blambda=\bmu$ when applicable.}}\\
%\cline{2-6}
%&\multicolumn{4}{c}{Excitations in the EM mode}&\\
&\centre{5}{$n_\textrm{exc}$}&\\
&0&1&2&3&4&&\\
\ns
$\tilde\mu$&\crule{5}&$\dim(M^{\bmu})$&$\dim(S^{\blambda})^{\rm a}$\\
&\centre{5}{$v_0^{(n_\textrm{exc},\tilde{\mu})}$}\\%{Excitations in the EM mode}\\
\mr%\hline
$0^N$&0&1&2&3&4&1&1\\
$0^{N-1}1^1$&&0&1&2&3&$N$&$N-1$\\
$0^{N-2}1^2$&&&0&1&2&$N(N-1)/2$&$N(N-3)/2$\\
$0^{N-1}2^1$&&&0&1&2&$N$&\\
$0^{N-3}1^3$&&&&0&1&$N(N-1)(N-2)/6$&$N(N-1)(N-5)/6$\\
$0^{N-2}1^1 2^1$&&&&0&1&$N(N-1)$&$(N-1)(N-2)/2$\\
$0^{N-1}3^1$&&&&0&1&$N$&\\
$0^{N-4}1^{4}$&&&&&0&$N(N-1)(N-2)(N-3)/24$&$N(N-1)(N-2)(N-7)/24$\\
$0^{N-3}1^{2}2^{1}$&&&&&0&$N(N-1)(N-2)/2$&$N(N-2)(N-4)/3$\\
$0^{N-2}2^{2}$&&&&&0&$N(N-1)/2$&\\
$0^{N-2}1^{1}3^{1}$&&&&&0&$N(N-1)$&\\
$0^{N-1}4^{1}$&&&&&0&$N$&\\
\cline{1-6}
%\rule{0pt}{12pt}
$\lvert\tilde\mu_{n_\textrm{exc}}\rvert$&1&2&4&7&12\\
%\rule{0pt}{12pt}
$\lvert{\lambda}_{n_\textrm{exc}}\rvert$&1&2&3&5&7\\
\br
\end{tabular*}
\noindent $^{\rm a}\blambda=\bmu$ when applicable.
%\end{indented}
%\endfulltable
\end{table}

%\begin{table}
%\centering
%\caption{Spectral configurations, EM mode excitations and degeneracies for the first five excitation manifolds\label{tab:specdeg}}
%\begin{indented}
%\item[]\begin{tabular}{*{15}{c}}
%$\tilde\mu$&$0^N$&$0^{N-1}1^1$&$0^{N-2}1^2$&$0^{N-1}2^1$&$0^{N-3}1^3$&$0^{N-2}1^1 2^1$&$0^{N-1}3^1$&$0^{N-4}1^4$&$0^{N-3}1^2 2^1$&$0^{N-2}2^2$&$0^{N-2}1^1 3^1$&$0^{N-1}4^1$&\multirow{2}{*}{$\lvert\tilde\mu_{n_\textrm{exc}}\rvert$}&\multirow{2}{*}{$\lvert{\lambda}_{n_\textrm{exc}}\rvert$}\\
%\cline{1-13}
%\rule{0pt}{12pt}{$n_\textrm{exc}$}&\multicolumn{12}{c}{Excitations in the EM mode}&\\
%\hline
%0&0&&&&&&&&&&&&1&1\\
%1&1&0&&&&&&&&&&&2&2\\
%2&2&1&0&0&&&&&&&&&4&3\\
%3&3&2&1&1&0&0&0&&&&&&7&5\\
%4&4&3&2&2&1&1&1&0&0&0&0&0&12&7\\
%
%\cline{1-13}
%\rule{0pt}{12pt}$\dim(M^{\bmu})$&1&$N$&$\binom{N}{2}$&$N$&$\binom{N}{3}$&$2\binom{N}{2}$&$N$&$\binom{N}{4}$&$3\binom{N}{3}$&$\binom{N}{2}$&$2\binom{N}{2}$&$N$&\\
%\rule{0pt}{12pt}$\dim(S^{\blambda})$\footnote{$\blambda=\bmu$ when applicable.}&1&$N-1$&$\frac{N}{2}(N-3)$&&$\frac{N-5}{N-2}\binom{N}{3}$&$\binom{N-1}{2}$&&$\frac{N-7}{N-3}\binom{N}{4}$&$\frac{2N}{N-3}\binom{N-2}{3}$&&&&
%\end{tabular}
%\end{indented}
%\end{table}

Since the emitters are considered as identical, $\hat{H}(N)$ is invariant under the action of the symmetric group of degree $N$, $S_N$, whose elements are the $N!$ possible permutations of labels of the emitters. In other words, for all permutations $\hat{\pi}\in S_N$, $\left[\hat{H}(N),\hat{\pi}\right]=0$.

The permutation operators act on the BEs according to
\begin{equation}
\hat{\pi}\ket{n_\textrm{exc},\tilde\mu;y}=\ket{n_\textrm{exc},\tilde\mu;y'},
\end{equation}
which means that BEs with common $n_\textrm{exc}$ and $\tilde\mu$ can be used to form a representation of $S_N$. For instance,
\begin{equation}\label{eq:s3perms}
\eqalign{
\hat{\pi}()=\ket{1_1}\bra{1_1}+\ket{1_2}\bra{1_2}+\ket{1_3}\bra{1_3},\\
\hat{\pi}(12)=\ket{1_1}\bra{1_2}+\ket{1_2}\bra{1_1}+\ket{1_3}\bra{1_3},\\
\hat{\pi}(23)=\ket{1_1}\bra{1_1}+\ket{1_2}\bra{1_3}+\ket{1_3}\bra{1_2},\\
\hat{\pi}(31)=\ket{1_1}\bra{1_3}+\ket{1_2}\bra{1_2}+\ket{1_3}\bra{1_1},\\
\hat{\pi}(123)=\ket{1_1}\bra{1_3}+\ket{1_2}\bra{1_1}+\ket{1_3}\bra{1_2},\\
\hat{\pi}(132)=\ket{1_1}\bra{1_2}+\ket{1_2}\bra{1_3}+\ket{1_3}\bra{1_1},\\}
\end{equation}
are the six elements of $S_3$ in the basis of PBEs with $\tilde\mu=1^1$.

The representations of $S_N$ identified with a partition $\bmu$ are, in general, reducible; they receive the name of \emph{permutation modules} and are denoted by $M^{\bmu}$ \cite{Meliot2017}. The dimension of this representation is the number of BEs with the same $\bmu$, i.e.,
\begin{equation}
\dim(M^{\bmu})=\frac{N!}{\displaystyle{\prod_{i=1}^{r+1}{\mu_{i}!}}}.
\end{equation}
%\begin{equation}\dim \left( {{M^{\bf{\mu }}}} \right) = \frac{{N!}}{{\prod\limits_{i = 1}^{{n_{\max }} + 1} {{\mu _i}!} }}\end{equation}
The permutation modules can, in turn, be decomposed in irreducible representations (irreps) according to Young's rule \cite{Sagan2013}:
\begin{equation}\label{eq:youngrule}
M^{\bmu}=\bigoplus_{\blambda\trianglerighteq\bmu}K_{\blambda\bmu} S^{\blambda},
\end{equation}
where $S^{\blambda}$ symbolizes the irreps, also known as \emph{Specht modules} \cite{Specht1935}.

The direct sum in  \eref{eq:youngrule} runs over all partitions $\blambda$ of $N$ that \emph{dominate} (denoted by the symbol $\trianglerighteq$) over $\bmu$, i.e., they fulfill \cite{Sagan2013}
\begin{equation}\label{eq:dominance}
\sum_{i=1}^j{\lambda_i}\geq\sum_{i=1}^j{\mu_i}\qquad\forall j: 1\leq j\leq r.
\end{equation}
For instance, consider the spectral configurations in the triply excited manifold: $0^N$, $0^{N-1}1^1$, $0^{N-2}1^2$, $0^{N-1}2^1$, $0^{N-3}1^3$, $0^{N-2}1^1 2^1$, and $0^{N-1}3^1$. Since they correspond to the same regular partition, the spectral configurations $0^{N-1}1^1$, $0^{N-1}2^1$ and $0^{N-1}3^1$  have identical permutational properties. Let us identify how each partition relates to $[N-2,1,1]$ in terms of dominance. Following  \eref{eq:dominance}, it is possible to conclude that
\begin{equation}
\eqalign{[N] \trianglerighteq [N-2,1,1],\cr
[N-1,1] \trianglerighteq [N-2,1,1],\cr
[N-2,2] \trianglerighteq [N-2,1,1],\cr
[N-3,3] \ntrianglerighteq [N-2,1,1],\cr
[N-2,1,1] \trianglerighteq [N-2,1,1].}
\end{equation}

The coefficients $K_{\blambda\bmu}$ in  \eref{eq:youngrule} are known as \emph{Kostka numbers} \cite{Kostka1882}; they indicate the number of times a permutation module contains a Specht module. While obtaining a closed  analytical expression to calculate them remains an open problem \cite{Lederer2005,Narayanan2006}, these coefficients can be found through their combinatorial interpretation: the number of semi-standard Young tableaux (SSYT) of shape $\blambda$ and content $\bmu$ \cite{Sagan2013}. A Young diagram (YD) of size $N$ and shape $\blambda$ is a collection of $N$ cells arranged in $r$ left-justified rows with $\lambda_i$ cells on the $i$th row. The present work uses the English notation, which is consistent with the regular form of the partitions (see \tref{tab:speconf} for selected examples). A SSYT of shape $\blambda$ and content $\bmu$ is obtained by filling in the cells of a $\blambda$-shaped YD with a collection of ordered symbols partitioned according to $\bmu$ in such a manner that the rows do not decrease to the right and the columns increase to the bottom \cite{Sagan2013}. For instance, the SSYT of shape $[N-1,1]$ and content $[N-2,1,1]$ are
\begin{equation}
\ytableausetup{boxsize=13pt}
\ytableaushort{00{\none[\dots]}01,2}\quad\textrm{and}\quad\ytableaushort{00{\none[\dots]}02,1};
\end{equation}
therefore, $K_{[N-1,1],[N-2,1,1]}=2$.

With the above definitions, it can be verified that the representation of $S_N$ in the basis of all BEs with ${\tilde\mu}=0^{N-2}1^1 2^1$ fulfills
\begin{equation}
M^{[N-2,1,1]}=S^{[N]}\oplus 2S^{[N-1,1]}\oplus S^{[N-2,2]}\oplus S^{[N-2,1,1]}.
\end{equation}

The dimension of the Specht modules corresponds to the number of standard Young tableaux (SYT) of shape $\blambda$, i.e., the number of ways in which the sequence $[1,2,\ldots,N]$ fills a $\blambda$-shaped YD such that the entries increase across rows and columns \cite{Sagan2013}. For instance, the SYTs for $\blambda=[3,2]$ are
\begin{equation}
\ytableaushort{123,45},~\ytableaushort{124,35},~\ytableaushort{125,34},~\ytableaushort{134,25},~\textrm{and}~\ytableaushort{135,24}.
\end{equation}
This quantity is given by the hook-length formula \cite{FrameRobinsonThrall1954}:
\begin{equation}\label{eq:hooklength}
\dim(S^{\blambda})=\frac{N!}{\displaystyle{\prod_{i=1}^{N}h_\lambda(i)}},%{\prod\limits_{i=1}^{N}h_\lambda(i)},
\end{equation}
where $h_\lambda(i)$ represents the number of cells in the hook of the $i$th cell in the $\blambda$-shaped YD, i.e. the number of cells that are either directly below of, directly to the right of, or the $i$th cell itself. For example, for ${\blambda}=[6,4,3,2,1]$, the hook of the cell with coordinates $(2,2)$ is
\begin{equation}
\ydiagram{6,1,3,1,1}*[*(gray)]{0,2+2,1+1,1+1}*[*(black)]{0,1+1},\nonumber
\end{equation}
and the hook-lengths for each cell are
\begin{equation}
\ytableaushort{{10}86421,7531,531,31,1}.\nonumber
\end{equation}
A direct result from Young's rule is that 
\begin{equation}
\dim(M^{\bmu})=\sum_{\blambda\trianglerighteq\bmu}K_{\blambda\bmu}\dim(S^{\blambda}).
\end{equation}
The dimensions of the permutation and Specht modules for the spectral configurations found up to $n_\textrm{exc}=4$ are displayed in \tref{tab:specdeg}.

The irreps identify the smallest possible subspaces that remain excluded under permutations. In other words, the space of BEs with the same $\tilde\mu$ can be split into sets of symmetry adapted linear combinations of BEs (SABEs) labeled by $\blambda$ for which
\begin{equation}\label{eq:sabedef}
\hat{\pi}\ket{n_\textrm{exc},{\tilde\mu},{\blambda};\mathbf{y},\mathbf{w}}=\ket{n_\textrm{exc},\tilde\mu,{\blambda};\mathbf{y}',\mathbf{w}},
\end{equation}
for all $\hat{\pi}\in S_n$, where the label $\mathbf{w}$ has been added to acknowledge repetition of irreps, i.e., when $K_{\blambda\bmu}>1$.

It is a well-known result from Representation Theory that
\begin{equation}\label{eq:schurweyl}
\left(\mathbb{C}^{r+1}\right)^{\otimes N}\cong \bigoplus_{\blambda\vdash N}\left(S^{\blambda}\otimes V^{\blambda}_{r+1}\right),
\end{equation}
 where $\mathbb{C}^{r+1}$ is the vector space spanned by the energy eigenstates of each molecule, and therefore $\left(\mathbb{C}^{r+1}\right)^{\otimes N}$ is the vector space spanned by the BEs. The symbol $\vdash$ reads as \emph{partition of}, and $V^{\blambda}_{r+1}$ is a so-called weight-space; it corresponds to an irrep of the unitary group $U(r+1)$. This result, known as the \emph{Schur-Weyl duality} \cite{Schur1901,Weyl2016}, implies that the SABEs are arranged in exclusive subspaces, labeled by $\blambda$, in which they are classified according to not only their behavior under permutations, but also under unitary operators. Furthermore, the decomposition in \eref{eq:schurweyl} provides with the meaning of all the indices in \eref{eq:sabedef}.

As previously discussed, the dimension of a Specht module corresponds with the number of SYT of shape $\blambda$, and the index $\mathbf{y}$ was used to enumerate them; therefore the indices $\blambda$ and $\mathbf{y}$ uniquely define a SYT. In what remains of this paper, the index $\mathbf{y}$ will encode the elements of the Young tableaux after removal of the top row. See some examples in \tref{tab:syt}.
\Table{Examples of standard Young tableaux and their associated indices for $N=6$. \label{tab:syt}}
\br
SYT&$\blambda$&$\mathbf{y}$\\
\mr
$\ytableaushort{123456}$&$[6]$&$0$\\
\ms
$\ytableaushort{12345,6}$&$[5,1]$&$6$\\
\ms
$\ytableaushort{1246,35}$&$[4,2]$&$(3,5)$\\
\ms
$\ytableaushort{124,356}$&$[3,3]$&$(3,5,6)$\\
\ms
$\ytableaushort{1256,3,4}$&$[4,1,1]$&$(3;4)$\\
\ms
$\ytableaushort{134,25,6}$&$[3,2,1]$&$(2,5;6)$\\
\br
\endtab
On the other hand, the dimension of the irreps of $U(r+1)$ correspond to the Kostka numbers. To be specific, the elements of the weight space $V^{\blambda}_{r+1}$ can be enumerated with SSYT, and thus uniquely identified with the indices $\tilde\mu$, $\blambda$, and $\mathbf{w}$. In the remaining of the present work, $\mathbf{w}$ will encode the elements of the SSYT after removal of the top row.  See some examples in \tref{tab:ssyt}.
\Table{\label{tab:ssyt}Examples of semistandard Young tableaux and their associated indices for $N=6$ and $\tilde\mu={0^3}{1^2}{2^1}$.}
\br
SSYT&$\blambda$&$\mathbf{w}$\\
\mr
$\ytableaushort{000112}$&$[6]$&$0$\\
\ms
$\ytableaushort{00012,1}$&$[5,1]$&$1^1$\\
\ms
$\ytableaushort{00011,2}$&$[5,1]$&$2^1$\\
\ms
$\ytableaushort{0001,12}$&$[4,2]$&${1^1}{2^1}$\\
\ms
$\ytableaushort{000,112}$&$[3,3]$&${1^2}{2^1}$\\
\ms
$\ytableaushort{0001,1,2}$&$[4,1,1]$&$(1^1,2^1)$\\
\ms
$\ytableaushort{000,11,2}$&$[3,2,1]$&$(1^2,2^1)$\\
\br
\endtab

To gain some insight into the meaning of the irreps, let us consider the global ground-state of the emitters, i.e., the state with $\tilde\mu=0^N$. The only partition that dominates over $[N]$ is itself, and there is a unique SSYT when the shape and content correspond to the same partition; therefore, $M^{[N]}=S^{[N]}$. These facts are consistent with the uniqueness of the state within each manifold where all the excitations reside in the EM mode, which is denoted by $\ket{n_\textrm{exc},0^N,[N];0,0}$.

\begin{figure*}
\tiny
\begin{center}
\begin{tikzpicture}[-latex ,auto]
\node (A) at (0,0) {$\ket{2,0^N,[N];0,0}$};
\node (A1) at (3.5,0) {$\ket{2,0^{N-1}1^1,[N];0,0}$};
\node (A2) at (7.5,0) {$\ket{2,0^{N-2}1^2,[N];0,0}$};
\node (B1) at (6.5,-1) {$\ket{2,0^{N-1}1^1,[N-1,1];k,{1^1}}$};
\node (B2) at (11.5,-1) {$\ket{2,0^{N-2}1^2,[N-1,1];k,{1^1}}$};
\node (C1) at (3.5,-2) {$\ket{2,0^{N-1}2^1,[N];0,0}$};
\node (C2) at (12.5,-2) {$\ket{2,0^{N-2}1^2,[N-2,2];(k',\ell),{1^2}}$};
\node (D) at (6.5,-3) {$\ket{2,0^{N-1}2^1,[N-1,1];k,{2^1}}$};
\path (A) edge node[above=-.1 cm]{$\hat{a}_0\hat{J}_{+}^{(1,0)}$} (A1);
\path (A1) edge node[above=-.1 cm] {$\hat{a}_0\hat{J}_{+}^{(1,0)}$} (A2);
\path (B1) edge node[above=-.1 cm] {$\hat{a}_0\hat{J}_{+}^{(1,0)}$} (B2);
\path (A1) edge node[right] {$\hat{a}_0\hat{J}_{+}^{(2,1)}$} (C1);
\path (B1) edge node[right] {$\hat{a}_0\hat{J}_{+}^{(2,1)}$} (D);
\path (A1) edge node[above right=-.1cm] {GS} (B1);
\path (C1) edge node[above right=-.1cm] {GS} (D);
\path (B2) edge node[above right=-.1cm] {GS} (C2);
\end{tikzpicture}
\end{center}
\caption{Diagram of the relations between SABEs in the doubly excited manifold. GS denotes Gram-Schmidt orthogonalization, $2\leq k\leq{N}$, $2\leq k'<\ell$, and $4\leq\ell\leq N$.\label{fig:2exman}}
\end{figure*}
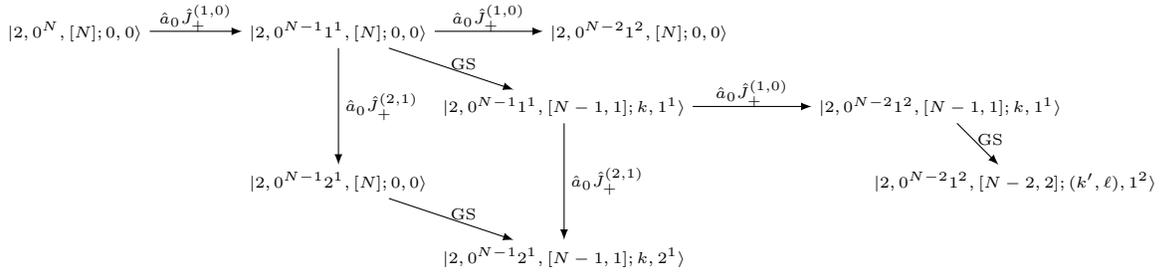

Since $\left[\hat{\pi},\hat{J}_{+}^{(u,v)}\right]=0$, the states $\ket{n_\textrm{exc},{\tilde\mu},{\blambda};\mathbf{y},\mathbf{w}}$ behave identically to $\hat{a}_0\hat{J}_{+}^{u,v}\ket{n_\textrm{exc},\tilde\mu,\blambda;\mathbf{y},\mathbf{w}}$ under permutations. This means that, even though the operator $\hat{a}_0\hat{J}_{+}^{(u,v)}$ modifies the spectral configuration, the states it couples carry the  same irrep. Consequently
\begin{equation}
\hat{a}_0\hat{J}_{+}^{(u,0)}\ket{n_\textrm{exc},0^N,[N];0,0}\propto\ket{n_\textrm{exc},0^{N-1}v^1,[N];0,0}.
\end{equation}
However, $M^{[N-1,1]}=S^{[N]}\oplus S^{[N-1,1]}$, which means that the remaining SABEs with $\tilde\mu=0^{N-1}u^1$ carry the irrep with $\blambda=[N-1,1]$. These states, $\ket{n_\textrm{exc},0^{N-1}v^1,[N-1,1];k,{1^1}}$ with $1\leq k\leq N-1$, can be obtained through Gram-Schmidt orthogonalization over the basis of PBEs but with $\ket{n_\textrm{exc},0^{N-1}v^1,[N];0,0}$ replacing one of the states and remaining fixed as seed of the procedure. The structure of the doubly excited manifold can be understood in terms of this procedure as illustrated in \fref{fig:2exman}. \Sref{sec:swbas} discusses the generalities of the expression of the SABEs in terms of the PBEs.

To exemplify how these concepts unfold, let us return to the representation in  \eref{eq:s3perms}. Starting from the state $\ket{1,0^3,[3];0,0}$, the action of the operator $\hat{a}_0\hat{J}_{+}^{(1,0)}$ yields
\begin{equation}
\ket{1,0^2 1^1,[3];0,0}=\frac{1}{\sqrt{3}}\left(\ket{1_1}+\ket{1_2}+\ket{1_3}\right).
\end{equation}
The remaining two SABEs with $\tilde\mu=0^2 1^1$ are
%%\numparts
%Fourier basis
%\begin{multline}
%\ket{1,0^2 1^1,[2,1];\pm1}=\\ \frac{e^{\pm\frac{2\pi i}{3}}\ket{1_1}+e^{\pm\frac{4\pi i}{3}}\ket{1_2}+e^{\pm2\pi i}\ket{1_3}}{\sqrt{3}},%\\
%%\ket{1,0^2 1^1,[2,1],1}=&\frac{e^{\frac{2\pi i}{3}}\ket{1_1}+e^{\frac{4\pi i}{3}}\ket{1_2}+e^{2\pi i}\ket{1_3}}{\sqrt{3}},
%\end{multline}
%%\endnumparts
%where $y$ has been chosen so it labels the wave vector. The permutation operators in this basis are
%\begin{widetext}
%\begin{equation}\label{eq:s3fourbas}
%\begin{split}
%\hat{\pi}()=&\ket{[3]}\bra{[3]}+\ket{[2,1];-1}\bra{[2,1];-1}+\ket{[2,1];1}\bra{[2,1];1},\\
%\hat{\pi}(12)=&\ket{[3]}\bra{[3]}+\ket{[2,1];-1}\bra{[2,1];1}+\ket{[2,1];1}\bra{[2,1];-1},\\
%\hat{\pi}(23)=&\ket{[3]}\bra{[3]}+e^{-\frac{2\pi i}{3}}\ket{[2,1];-1}\bra{[2,1];1}+e^{\frac{2\pi i}{3}}\ket{[2,1];1}\bra{[2,1];-1},\\
%\hat{\pi}(31)=&\ket{[3]}\bra{[3]}+e^{\frac{2\pi i}{3}}\ket{[2,1];-1}\bra{[2,1];1}+e^{-\frac{2\pi i}{3}}\ket{[2,1];1}\bra{[2,1];-1},\\
%\hat{\pi}(123)=&\ket{[3]}\bra{[3]}+e^{\frac{2\pi i}{3}}\ket{[2,1];-1}\bra{[2,1];-1}+e^{-\frac{2\pi i}{3}}\ket{[2,1];1}\bra{[2,1];1},\\
%\hat{\pi}(132)=&\ket{[3]}\bra{[3]}+e^{-\frac{2\pi i}{3}}\ket{[2,1];-1}\bra{[2,1];-1}+e^{\frac{2\pi i}{3}}\ket{[2,1];1}\bra{[2,1];1},
%\end{split}
%\end{equation}
%\end{widetext}
%where the notation was simplified by making implicit the common indices $n_\textrm{exc}=1$ and $\tilde\mu=0^{2}1^{1}$%, as well as by writing $\ket{[3]}$ instead of $\ket{[3];1}$
\numparts
\begin{equation}
\ket{1,0^2 1^1,[2,1];2,{1^1}}=\frac{\ket{1_1}-\ket{1_2}}{\sqrt{2}},
\end{equation}
and
\begin{equation}
\ket{1,0^2 1^1,[2,1];3,{1^1}}=\frac{-\ket{1_1}-\ket{1_2}+2\ket{1_3}}{\sqrt{6}}.
\end{equation}
\endnumparts
The permutation operators in this basis are

\begin{equation}\label{eq:s3mlbas}
\eqalign{
\fl\hat{\pi}()=\ket{[3];0}\bra{[3];0}+\ket{[2,1];2}\bra{[2,1];2}+\ket{[2,1];3}\bra{[2,1];3},\\
\fl\hat{\pi}(12)=\ket{[3];0}\bra{[3];0}-\ket{[2,1];2}\bra{[2,1];2}+\ket{[2,1];3}\bra{[2,1];3},\\
\fl\hat{\pi}(23)=\ket{[3];0}\bra{[3];0}+\tfrac{1}{2}\left(\ket{[2,1];2}\bra{[2,1];2}-\ket{[2,1];3}\bra{[2,1];3}\right)\\
+\tfrac{\sqrt{3}}{2}\left(\ket{[2,1];2}\bra{[2,1];3}+\ket{[2,1];3}\bra{[2,1];2}\right),\\
\fl\hat{\pi}(31)=\ket{[3];0}\bra{[3];0}+\tfrac{1}{2}\left(\ket{[2,1];2}\bra{[2,1];2}-\ket{[2,1];3}\bra{[2,1];3}\right)\\
-\tfrac{\sqrt{3}}{2}\left(\ket{[2,1];2}\bra{[2,1];3}+\ket{[2,1];3}\bra{[2,1];2}\right),\\
\fl\hat{\pi}(123)=\ket{[3];0}\bra{[3];0}-\tfrac{1}{2}\left(\ket{[2,1];2}\bra{[2,1];2}+\ket{[2,1];3}\bra{[2,1];3}\right)\\
-\tfrac{\sqrt{3}}{2}\left(\ket{[2,1];2}\bra{[2,1];3}-\ket{[2,1];3}\bra{[2,1];2}\right),\\
\fl\hat{\pi}(132)=\ket{[3];0}\bra{[3];0}-\tfrac{1}{2}\left(\ket{[2,1];2}\bra{[2,1];2}+\ket{[2,1];3}\bra{[2,1];3}\right)\\
+\tfrac{\sqrt{3}}{2}\left(\ket{[2,1];2}\bra{[2,1];3}-\ket{[2,1];3}\bra{[2,1];2}\right).
}
\end{equation}
where the notation was simplified by making implicit the common indices $n_\textrm{exc}=1$ and $\tilde\mu=0^{2}1^{1}$, as well as by omitting the label $\mathbf{w}$. As it can be seen in  \eref{eq:s3mlbas}, the permutation operators have two clear independent subspaces labeled by $[3]$ and $[2,1]$, respectively.

%As stated before, two spectral configurations display the same permutational features as long as they correspond to the same partition, e.g., $0^{N-4}1^{2}2^{1}3^{1}$ and $1^{1}2^{N-4}4^{1}5^{2}$. Since all spectral configurations can be constructed as excitations from the groundstate, and excited states ``inherit'' its symmetry from lower energy states, the symmetry label $\lambda$ only features spectral configurations that correspond to so-called \emph{partitions with weakly decreasing multiplicities}, i.e., for a given set of populations $\{n_\lambda(v)\}$, the only populated levels $\{v\}$ are those which minimize the quantity $\prod_{v=1}^{r}v^{n_\lambda(v)}$. For instance, while $\mu=1^{1}2^{N-4}4^{1}5^{2}$ is a perfectly valid spectral configuration, it cannot act as a symmetry label; instead, $\lambda=0^{N-4}1^{2}2^{1}3^{1}$ is the symmetry label with the same partition.
The spectral configurations, $\tilde{\mu}$, giving rise to partitions that are valid labels, $\blambda$, for the irreps fulfill
\begin{equation}
n_{\tilde{\mu}}(v)\geq n_{\tilde{\mu}}(v')\qquad\forall v,v':v'>v>0;
\end{equation}
this constraint produces the so-called partitions with weakly decreasing multiplicities \cite{Sloane}. The enumeration of these partitions imply that the total number of irreps in the $n_\textrm{exc}$th manifold is
\begin{equation}\label{eq:sympart}
\lvert\lambda_{n_\textrm{exc}}\rvert=\sum_{n=0}^{n_\textrm{exc}}q(n),
\end{equation}
where $q(n)$ is the number of partitions of $n$ with weakly decreasing multiplicities as illustrated in \tref{tab:specdeg}. These numbers can be extracted from the expansion \cite{AbramowitzStegun1970,Andrews1998}
\begin{equation}\label{eq:sympartseries}
\prod_{k=2}^{m}{\left(1-x^{\binom{k}{2}}\right)}^{-1}=\sum_{n=0}^m{q(n)x^n}+\Or\left(x^{m+1}\right).
\end{equation}

The main result of this section is that the Hamiltonians of the excitation manifolds can be split according to
\begin{equation}
\hat{H}_{n_\textrm{exc}}(N)=\sum_{i=1}^{\left\lvert\lambda_{n_\textrm{exc}}\right\rvert}{\hat{H}_{n_\textrm{exc}}^{(\blambda_i)}},\label{eq:nexcsplit}
\end{equation}
where
\begin{equation}
\hat{H}_{n_\textrm{exc}}^{(\blambda)}=\sum_{j=1}^{\dim\left(S^{\blambda}\right)}{\hat{H}_{n_\textrm{exc}}^{\left(\blambda,\mathbf{y}_j\right)}},\label{eq:lambdasplit}
\end{equation}
with each $\hat{H}_{n_\textrm{exc}}^{(\blambda,\mathbf{y})}$ encoding the energies and interactions of spectral configurations with the same symmetry. Equation \eref{eq:lambdasplit} implies that the states obtained from diagonalizing $\hat{H}_{n_\textrm{exc}}^{(\blambda)}$ are $\dim(S^{\blambda})$-fold degenerate. In particular, for the diagonal part of the Hamiltonian, we have

\begin{equation}
%\begin{split}
\hat{H}_{\textrm{bare},n_\textrm{exc}}^{(\blambda)}=\sum_{\tilde\mu:\bmu\trianglelefteq\blambda}\sum_{i=1}^{\dim(S^{\blambda})}\sum_{j=1}^{K_{\boldsymbol{\lambda\mu}}}\varepsilon_{n_\textrm{exc}}^{(\tilde\mu)}\ket{n_\textrm{exc},\tilde\mu,\blambda;\mathbf{y}_i,\mathbf{w}_j}\bra{n_\textrm{exc},\tilde\mu,\blambda;\mathbf{y}_i,\mathbf{w}_j},%\\
%=&\left(\sum_{\tilde\mu:\bmu\trianglelefteq\blambda}\sum_{ t=1}^{K_{\boldsymbol{\lambda\mu}}}\varepsilon_{n_\textrm{exc}}^{(\tilde\mu)}\ket{n_\textrm{exc},\tilde\mu,\blambda;1, t}\bra{n_\textrm{exc},\tilde\mu,\blambda;1, t}\right)\otimes\mathbf{1}_{\dim(S^{\blambda})},
%\textrm{diag}\left(\hat{H}_3^{[N]}\right)=&
%\begin{pmatrix}
%\varepsilon_3^{(0^N)}&\varepsilon_3^{(0^{N-1}1^{1})}&\varepsilon_3^{(0^{N-2}1^{2})}&\varepsilon_3^{(0^{N-1}2^1)}&\varepsilon_3^{(0^{N-3}1^{3})}&\varepsilon_3^{(0^{N-2}1^{1}2^{1})}&\varepsilon_3^{(0^{N-1}3^1)}
%\end{pmatrix}\\
%\textrm{diag}\left(\hat{H}_3^{[N-1,1]}\right)=&
%\begin{pmatrix}
%\varepsilon_3^{(0^{N-1}1^1)}&\varepsilon_3^{(0^{N-2}1^2)}&\varepsilon_3^{(0^{N-1}2^1)}&\varepsilon_3^{(0^{N-3}1^{3})}&\varepsilon_3^{(0^{N-2}1^{1}2^{1})}&\varepsilon_3^{(0^{N-2}1^{1}2^{1})}&\varepsilon_3^{(0^{N-1}3^1)}
%\end{pmatrix}
%\otimes \mathbf{1}_{N-1},\\
%\textrm{diag}\left(\hat{H}_3^{[N-2,2]}\right)=&
%\begin{pmatrix}
%\varepsilon_3^{(0^{N-2}1^2)}&\varepsilon_3^{(0^{N-3}1^{3})}&\varepsilon_3^{(0^{N-2}1^{1}2^{1})}
%\end{pmatrix}
%\otimes \mathbf{1}_{N(N-3)/2},\\
%\textrm{diag}\left(\hat{H}_3^{[N-3,3]}\right)=&
%\begin{pmatrix}
%\varepsilon_3^{(0^{N-3}1^{3})}
%\end{pmatrix}
%\otimes \mathbf{1}_{N(N-1)(N-5)/6},\\
%\textrm{diag}\left(\hat{H}_3^{[N-2,1,1]}\right)=&
%\begin{pmatrix}
%\varepsilon_3^{(0^{N-2}1^{1}2^{1})}
%\end{pmatrix}
%\otimes \mathbf{1}_{(N-1)(N-2)/2},
%\end{split}
\end{equation}

%where $\mathbf{1}_n$ is the $n\times n$ identity matrix. 
Notice that
\begin{equation}
\dim\left(\hat{H}_{n_\textrm{exc}}^{(\blambda,\mathbf{y})}\right)=\sum_{\bmu\trianglelefteq\blambda}K_{\blambda\bmu}.
\end{equation}
The simplification achieved with this strategy is significant since, for each excitation manifold, the problem has been reduced to the diagonalization of $\lvert\lambda_{n_\textrm{exc}}\rvert\sim \exp(A n_\textrm{exc}^{1/3})$ matrices with dimensions in the neighborhood of $\lvert\tilde\mu_{n_\textrm{exc}}\rvert\sim\exp(\sqrt{C n_\textrm{exc}})$ \cite{Almkvist2006}, both independent of $N$, which is a substantial gain over the original $\binom{N+n_\textrm{exc}}{n_\textrm{exc}}$-dimensional matrices.

A complete discussion of the structure of the Hamiltonians requires to include the couplings between spectral configurations induced by collective excitations; these will be explored in \sref{sec:colcoupl}.

\subsection{Schur-Weyl basis.\label{sec:swbas}}

The SABEs obtained through application of the collective excitations, $\hat{J}_{+}^{(u,v)}$, and Gram-Schmidt orthogonalization have been discussed in the literature under the name of Schur-Weyl states \cite{HaaseButler1984,JakubczykLulekJakubczykEtAl2010,JakubczykKravetsJakubczyk2011,BoteroMejia2018}, or Gelfand-Tsetlin states \cite{Gould1989,Molev1994,CorderoCastanosLopez-PenaEtAl2013,FutornyRamirezZhang2019}. Although not necessary for the block-diagonalization of the Hamiltonian, an illustration of the explicit form of the symmetrized states in terms of the PBEs might be useful for the reader. Since symmetrization only affects the portion of the Hilbert space concerning the emitters, the quantum number $n_\textrm{exc}$, as well as the operators acting on the photonic mode, $\hat{a}_0^{(\dagger)}$, are not included in the following discussion, i.e., the SABEs are denoted by $\ket{\tilde{\mu},\blambda;\mathbf{y},\mathbf{w}}$.

If the explicit form of the SABE in terms of PBEs is known, the application of collective excitation operators gives a straightforward result. For instance, the ground-state and some of its collective excitations are
\numparts
\begin{eqnarray}
\ket{0^N,[N];0,0}=\ket{0},\\
\ket{0^{N-1}v^{1},[N];0,0}=\frac{1}{\sqrt{N}}\sum_{i=1}^N\ket{v_i},\\
\ket{0^{N-2}v^{2},[N];0,0}=\binom{N}{2}^{-1/2}\sum_{i=1}^{N-1}\sum_{j=i+1}^{N}\ket{v_i v_j},\\
\ket{0^{N-3}v^{3},[N];0,0}=\binom{N}{3}^{-1/2}\sum_{i=1}^{N-2}\sum_{j=i+1}^{N-1}\sum_{k=j+1}^{N}\ket{v_i v_j v_k},\\
\ket{0^{N-2}u^{1}v^{1},[N];0,0}=\frac{1}{\sqrt{N(N-1)}}\sum_{i=1}^{N-1}\sum_{j=i+1}^{N}\left(\ket{u_i v_j}+\ket{v_i u_j}\right),
\end{eqnarray}
\endnumparts
where $\blambda=[N]$ and $\mathbf{y}=0$ indicate the SYT $\ytableaushort{12{\none[\dots]}N}$. This tableau implies that the wavefunction must be invariant under any permutation of all the labels. Since all emitters are in the same state in the collective ground-state, this condition is met by default. The linear combinations of the excited spectral configurations are thus totally symmetric.

The states with spectral configuration such that $\bmu=\blambda$ cannot be generated through excitation operators, and thus require a orthogonalization strategy. Since these states are highly degenerate, the change of basis is not unique; however, there is an approach that highlights the meaning of the label $\mathbf{y}$. For a 1D array $(i_1,i_2,\dots,i_n)$, let's define the vandermonde matrix $\hat{\mathbf{V}}(i_1,i_2,\dots,i_n)$ with elements
\begin{equation}
\left[\hat{\mathbf{V}}(i_1,i_2,\dots,i_n)\right]_{\alpha,\beta}=\hat{\sigma}_{\alpha-1,0}^{(\beta)}.
\end{equation}
A given SYT can be associated to Young operators of the form
\begin{equation}
\hat{Y}(\blambda,\mathbf{y})=\sum_{\psi}\prod_{j=1}^{\lambda_1}\det\left(\hat{\mathbf{V}}(\textrm{col}_j[\mathbf{y}'_\psi])\right),
\end{equation}
where $\textrm{col}_j(\mathbf{A})$ extracts the $j$th column of array $\mathbf{A}$, and $\mathbf{y}'_\psi$ is any permutation of the elements in each row of $\mathbf{y}$ that produces downwards increasing columns. For instance, the SYT
\begin{equation}
\mathbf{y}'_1=\ytableaushort{134,25,6},
\end{equation}
generates the arrays
\numparts
\begin{eqnarray}
\mathbf{y}'_2=\ytableaushort{134,2{\none}5,6},\\
\mathbf{y}'_3=\ytableaushort{134,25,{\none}6},\\
\mathbf{y}'_4=\ytableaushort{134,2{\none}5,{\none}{\none}6}.
\end{eqnarray}
\endnumparts
And the corresponding operator is
\begin{equation}
\eqalign{
\fl\hat{Y}\left([3,2,1],(2,5;6)\right)=\det\left(\hat{\mathbf{V}}(1,2,6)\right)\det\left(\hat{\mathbf{V}}(3,5)\right)\det\left(\hat{\mathbf{V}}(4)\right)\\
+\det\left(\hat{\mathbf{V}}(1,2,6)\right)\det\left(\hat{\mathbf{V}}(3)\right)\det\left(\hat{\mathbf{V}}(4,5)\right)\\
+\det\left(\hat{\mathbf{V}}(1,2)\right)\det\left(\hat{\mathbf{V}}(3,5,6)\right)\det\left(\hat{\mathbf{V}}(4)\right)\\
+\det\left(\hat{\mathbf{V}}(1,2)\right)\det\left(\hat{\mathbf{V}}(3)\right)\det\left(\hat{\mathbf{V}}(4,5,6)\right)
}
\end{equation}
If $\bmu=\blambda$, the wavefunction $\ket{\tilde{\mu},\blambda;\mathbf{y},\mathbf{w}}$ is proportional to $\hat{Y}(\blambda,\mathbf{y})\ket{0^N,[N];0,0}$. Notice that the constraint $\bmu=\blambda$ uniquely defines a SSYT, and therefore a $\mathbf{w}$.

The states orthogonal to $\ket{0^{N-1}1^{1},[N];0,0}$ with the same $\tilde{\mu}$ can be written as $\ket{0^{N-1}1^{1},[N-1,1];k,1^1}$, with $2\leq k\leq N$. The corresponding SYT are 
\begin{equation}
\ytableausetup
{mathmode, boxsize=2em}
\begin{ytableau}
1&2&\dots&\scriptstyle k-1&\scriptstyle k+1&\dots&N\\
k
\end{ytableau},\nonumber
\end{equation}
which have associated Young operators of the form
\begin{equation}
\hat{Y}([N-1,1],k)=\sum_{i=1}^{k-1}\prod_{j=1}^{i-1}\hat{\sigma}_{0,0}^{(j)}\cdot
\begin{vmatrix}
\hat{\sigma}_{0,0}^{(i)}&\hat{\sigma}_{0,0}^{(k)}\\
\hat{\sigma}_{1,0}^{(i)}&\hat{\sigma}_{1,0}^{(k)}
\end{vmatrix}
\cdot\prod_{j'=i+1}^{N}\hat{\sigma}_{0,0}^{(j')}.
\end{equation}
Therefore, the corresponding SABEs are
\begin{equation}
\ket{0^{N-1}1^{1},[N-1,1];k,1^1}=\frac{1}{\sqrt{k(k-1)}}\left(-\sum_{i=1}^{k-1}\ket{1_i}+
(k-1)\ket{1_k}\right).
\end{equation}
The functions $\ket{0^{N-1}v^{1},[N-1,1];k,v^1}$ for $v>1$ have an equivalent form. Furthermore, the action of the collective excitation $\hat{J}_{+}^{(1,0)}$ yields
\begin{equation}
\eqalign{
\fl\ket{0^{N-2}1^{2},[N-1,1];k,1^1}=\frac{1}{{\sqrt {\left( {N - 2} \right)k\left( {k - 1} \right)} }}\left\{ \sum_{i = 1}^{k - 1} {\left[ \left( {k - 2} \right)\ket {{1_i}{1_k}}-\sum_{j = k + 1}^N {\ket {{1_i}{1_j}} } \right]}\right.\\
\left. +\left( {k - 1} \right)\sum_{i = k + 1}^N \ket{{1_k}{1_i}}-2\sum_{i = 1}^{k - 2} {\sum_{j = i + 1}^{k - 1} {\ket {{1_i}{1_j}}} }  \right\}
}
\end{equation}

Following the same reasoning, the remaining states with $\tilde{\mu}={0^{N-2}}{1^2}$ that are orthogonal to all the states above are
\begin{equation}
\eqalign{
\fl\ket{0^{N-2}1^{2},[N-2,2];(k,\ell),1^2}=\frac{1}{\sqrt{k(k-1)(\ell-2)(\ell-3)}}\left\{2\sum\limits_{i=1}^{k-2}\sum\limits_{j=i+1}^{k-1}\ket{1_i 1_j}\right.\\
-(k-2)\sum\limits_{i=1}^{k-1}\ket{1_i 1_k}-\sum\limits_{j=k+1}^{\ell-1}\left[(k-1)\ket{1_k 1_j}-\sum\limits_{i=1}^{k-1}\ket{1_i 1_j}\right]\\
\left.+(\ell-3)\left[(k-1)\ket{1_k 1_\ell}-\sum\limits_{i=1}^{k-1}\ket{1_i 1_\ell}\right]\right\}
}
\end{equation}

The operators that act on the ground-state to generate all the SABEs relevant up to the triply excited manifold are tabulated in \ref{sec:swbasoper}.

\section{Collective couplings.\label{sec:colcoupl}}

After determining the SABEs, it is possible to evaluate the matrix elements of the interaction Hamiltonian. Let's consider the collective transition of a quantum from level $v^{*}$ to $v^{*}+s$, i.e., the action of the operator $\hat{J}_{+}^{v^{*}+s,v^{*}}$ on a SABE with spectral configuration $\tilde{\mu}$. As a result of the transition, the new SABE belongs to the spectral configuration $\tilde{\mu}'$, which relates to the initial spectral configuration according to
\begin{equation}
s=\sum_{v=1}^{n_\textrm{exc}}v[n_{\tilde\mu'}(v)-n_{\tilde\mu}(v)].
\end{equation}
Additionally, the occupation of levels in the two spectral configurations fulfill
\begin{equation}
n_{\tilde{\mu}}(v^{*})-n_{\tilde{\mu}'}(v^{*})=n_{\tilde{\mu}'}(v^{*}+s)-n_{\tilde{\mu}}(v^{*}+s)=1.
\end{equation}

Given the definition of $\hat{H}_\textrm{int}(N)$ in \eref{eq:hamint}, the only possible transitions are those between neighboring levels; consequentially, $s=\pm1$. It can be shown that under this constraint, for a given manifold, the number of pairings $\{\tilde\mu',\tilde\mu\}$ afforded by photonic (de)excitations is
\begin{equation}\label{eq:numpairs}
n_\textrm{int}(n_\textrm{exc})=\sum_{i=1}^{n_\textrm{exc}-1}\binom{n_\textrm{exc}-i+2}{i+1}.
\end{equation}

The matrix elements of the interaction Hamiltonian are
\begin{equation}\label{eq:offdiag}
\eqalign{\fl\bra{n'_\textrm{exc},\tilde\mu',\blambda';\mathbf{y}',\mathbf{w}'}\hat{H}_{\textrm{int}}(N)\ket{n_\textrm{exc},\tilde\mu,\blambda;\mathbf{y},\mathbf{w}}\\
=g_{\tilde\mu',\tilde\mu}B^{(n_\textrm{exc})}_{\tilde\mu',\tilde\mu}L_{\tilde\mu',\tilde\mu}^{(\blambda)}C_{\tilde\mu',\tilde\mu}^{(\blambda)}(\mathbf{w}',\mathbf{w})\delta_{n'_\textrm{exc}n_\textrm{exc}}\delta_{\blambda'\blambda}\delta_{\mathbf{y}'\mathbf{y}},}
\end{equation}
where, in agreement with the discussion above,
\begin{equation}\label{eq:coupconst}
g_{\tilde\mu',\tilde\mu}=g_{v^*+s,v^*}\delta_{\lvert s\rvert,1}\delta_{n_{\tilde\mu'}(v^*),n_{\tilde\mu}(v^*)-1}\delta_{n_{\tilde\mu'}(v^*+s),n_{\tilde\mu}(v^*+s)+1},
\end{equation}
while the coefficient 
\begin{equation}\label{eq:couphot}
B^{(n_\textrm{exc})}_{\tilde\mu',\tilde\mu}=\left(v_0^{(n_\textrm{exc},\tilde\mu)}+\delta_{s,-1}\right)^{1/2}
\end{equation}
accounts for the (de)excitation of the EM mode.

To describe the redistribution of quanta among the emitters, let's define the operators
\begin{equation}
\hat{J}_r^{(\tilde\mu',\tilde\mu)}=\hat{J}_r^{(v^*+s,v^*)},
\end{equation}
where $r\in\{-,0,+\}$, that generate the $\mathfrak{su}(2)$ algebra with Casimir operator
\begin{equation}
\hat{\mathbf J}^2_{\tilde\mu',\tilde\mu}=\left(\hat{J}_0^{(\tilde\mu',\tilde\mu)}\right)^2+\tfrac{1}{2}\left(\hat{J}_{+}^{(\tilde\mu',\tilde\mu)}\hat{J}_{-}^{(\tilde\mu',\tilde\mu)}+\hat{J}_{-}^{(\tilde\mu',\tilde\mu)}\hat{J}_{+}^{(\tilde\mu',\tilde\mu)}\right).
\end{equation}
These operators allow the definition of the quantities
\begin{equation}
M_{\tilde\mu',\tilde\mu}=\bra{\tilde\mu}\hat{J}_0^{(\tilde\mu',\tilde\mu)}\ket{\tilde\mu}
\end{equation}
and
\begin{equation}
J_{\tilde\mu',\tilde\mu}^{(\blambda)}\left(J_{\tilde\mu',\tilde\mu}^{(\blambda)}+1\right)=\bra{\tilde\mu,\blambda}\hat{\mathbf J}^2_{\tilde\mu',\tilde\mu}\ket{\tilde\mu,\blambda},
\end{equation}
where the explicit labels in bras and kets are the only relevant ones in the calculation of the indicated matrix elements as pointed out in the study of two-level systems, where the SABEs are known as Dicke states, and $J_{\tilde\mu',\tilde\mu}^{(\blambda)}$ as the Dicke cooperation number \cite{Dicke1954}. To be specific
\begin{equation}
M_{\tilde\mu',\tilde\mu}=\frac{n_{\tilde\mu}(v^*+s)-n_{\tilde\mu}(v^*)}{2},\label{eq:angmomm}
\end{equation}
and
\begin{equation}
J_{\tilde\mu',\tilde\mu}^{(\blambda)}={\max_{\tilde\mu''}}\left(\bra{\tilde\mu'',\blambda}\hat{J}_0^{(\tilde\mu',\tilde\mu)}\ket{\tilde\mu'',\blambda}\right)\label{eq:angmomj}.
\end{equation}

According to the prescription of quantum angular momentum, the contribution of collective (de)excitations to the coupling between states is given by
\begin{equation}\label{eq:angmomcoup}
L_{\tilde\mu',\tilde\mu}^{(\blambda)}=\left[\left(J_{\tilde\mu',\tilde\mu}^{(\blambda)}+M_{\tilde\mu',\tilde\mu}\right)\left(J_{\tilde\mu',\tilde\mu}^{(\blambda)}-M_{\tilde\mu,\tilde\mu'}\right)\right]^{1/2}.
\end{equation}
Couplings are said to be \emph{superradiant} if $L_{\tilde\mu',\tilde\mu}^{(\blambda)}>1$, and \emph{subradiant} if $L_{\tilde\mu',\tilde\mu}^{(\blambda)}<1$ \cite{Dicke1954}. \ref{ap:calcL} shows a couple of neat ways to calculate these coefficients.

The last component of  \eref{eq:offdiag}, the coefficient $C_{\tilde\mu',\tilde\mu}^{(\blambda)}(\mathbf{w}',\mathbf{w})$, acknowledges the fact that, since $\left[\hat{J}_{+}^{(v',v)},\hat{J}_{+}^{(u',u)}\right]\propto1-\delta_{uv}\delta_{u'v'}$, two strings of the same raising operators applied in different order to a SABE will not, in general, produce the same state. This situation creates an ambiguity when $\hat{H}_{n_\textrm{exc}}^{(\blambda,\mathbf{y})}$ requires more than one function with the same $\tilde\mu$, i,e, when $K_{\blambda\bmu}>1$; otherwise, $C_{\tilde\mu',\tilde\mu}^{(\blambda)}(\mathbf{w}',\mathbf{w})=1$. For instance, consider the state $\ket{3,0^{N-1}1^{1},[N-1,1];2,1}=(\ket{2_{0}1_{1}}-\ket{2_{0}1_{2}})/\sqrt{2}$, and apply to it consecutive excitations to get to a state with $\tilde\mu=0^{N-2}1^{1}2^{1}$. There are two possible paths:

\numparts
\begin{equation}
\eqalign{\fl\frac{\ket{2_{0}1_{1}}-\ket{2_{0}1_{2}}}{\sqrt{2}}\xrightarrow{\hat{a}_0\hat{J}_{+}^{(1,0)}}\sum_{i=3}^N\frac{\ket{1_{0}1_{1}1_{i}}-\ket{1_{0}1_{2}1_{i}}}{\sqrt{2(N-2)}}\\
\xrightarrow{\hat{a}_0\hat{J}_{+}^{(2,1)}}\sum_{i=3}^N\frac{\ket{2_{1}1_{i}}-\ket{2_{2}1_{i}}+\ket{1_{1}2_{i}}-\ket{1_{2}2_{i}}}{2\sqrt{N-2}},}
\end{equation}
and

\begin{equation}
\eqalign{\fl\frac{\ket{2_{0}1_{1}}-\ket{2_{0}1_{2}}}{\sqrt{2}}\xrightarrow{\hat{a}_0\hat{J}_{+}^{(2,1)}}\frac{\ket{1_{0}2_{1}}-\ket{1_{0}2_{2}}}{\sqrt{2}}\\
\xrightarrow{\hat{a}_0\hat{J}_{+}^{(1,0)}}\frac{1}{\sqrt{2(N-1)}}\left[\ket{2_{1}1_{2}}-\ket{1_{1}2_{2}}+\sum_{i=3}^N(\ket{2_{1}1_{i}}-\ket{2_{2}1_{i}})\right].}
\end{equation}
\endnumparts

While the identifications
\begin{equation}
\eqalign{
\ket{3,0^{N-2}1^{2},[N-1,1];2,1}=\sum_{i=3}^N\frac{\ket{1_{0}1_{1}1_{i}}-\ket{1_{0}1_{2}1_{i}}}{\sqrt{2(N-2)}},\\
\ket{3,0^{N-1}2^{1},[N-1,1];2,1}=\frac{\ket{1_{0}2_{1}}-\ket{1_{0}2_{2}}}{\sqrt{2}}
}
\end{equation}
 are immediate, the assignment of $\ket{3,0^{N-2}1^{1}2^{1},[N-1,1];2,1}$ and $\ket{3,0^{N-2}1^{1}2^{1},[N-1,1];2,2}$ is unclear since the two obtained states are linearly independent yet not mutually orthogonal. Bypassing this ambiguity requires the definition of an orthonormal basis spanned by the previous states, then the coefficients can be extracted from the identity
\begin{equation}\label{eq:lccoup}
L_{\tilde\mu'\tilde\mu}^{(\blambda)}C_{\tilde\mu'\tilde\mu}^{(\blambda)}(\mathbf{w}',\mathbf{w})=\bra{\tilde\mu,\blambda;\mathbf{y},\mathbf{w}}\hat{J}_{-}^{(\tilde\mu',\tilde\mu)}\ket{\tilde\mu',\blambda;\mathbf{y},\mathbf{w}'}.
\end{equation}

For instance, let's set
\begin{equation}
\ket{A}=\sum_{i=3}^N\frac{\ket{2_{1}1_{i}}-\ket{2_{2}1_{i}}+\ket{1_{1}2_{i}}-\ket{1_{2}2_{i}}}{2\sqrt{N-2}},
\end{equation}
and
\begin{equation}
\ket{X}=\frac{\ket{2_{1}1_{2}}-\ket{1_{1}2_{2}}+\sum_{i=3}^N(\ket{2_{1}1_{i}}-\ket{2_{2}1_{i}})}{\sqrt{2(N-1)}},
\end{equation}
and compute their orthogonal complements:
\begin{equation}
\ket{B}=\frac{\hat{I}-\ket{A}\bra{A}}{\left(1-\lvert\braket{X}{A}\rvert^2\right)^{1/2}}\ket{X},
\end{equation}
and
\begin{equation}
\ket{Y}=\frac{\hat{I}-\ket{X}\bra{X}}{\left(1-\lvert\braket{X}{A}\rvert^2\right)^{1/2}}\ket{A},
\end{equation}
where $\hat{I}$ is the identity operator. The values of the coefficients calculated with both working basis are shown in \tref{tab:cproj}. Since the bare energies, $\varepsilon_{n_\textrm{exc}}^{(\tilde\mu)}$, are independent of the index $\mathbf{w}$, and $\sum_{ t'=1}^{K_{\boldsymbol{\lambda\mu}'}}\left\lvert C_{\tilde\mu'\tilde\mu}^{(\blambda)}(\mathbf{w}',\mathbf{w})\right\rvert^2=1$, the choice of basis is immaterial for the eigenvalues of $\hat{H}_{n_\textrm{exc}}^{(\blambda,\mathbf{y})}$.
\Table{\label{tab:cproj}Coefficients of projected couplings involving $\tilde\mu'=0^{N-2}1^{1}2^{1}$ for repeated irreps with $\blambda=[N-1,1]$.}
\br
$\ket{3,\tilde\mu',\blambda;2,1}$&$\ket{3,\tilde\mu',\blambda;2,2}$&$\tilde\mu$&$C_{\tilde\mu',\tilde\mu}^{(\blambda)}({1^1},{1^1})$&$C_{\tilde\mu',\tilde\mu}^{(\blambda)}({1^1},{2^1})$\\
\mr
{$\ket{A}$}&{$\ket{B}$}&$0^{N-2}1^{2}$&1&0\\
&&$0^{N-1}2^{1}$&$\braket{X}{A}$&$\braket{X}{B}$\\
{$\ket{X}$}&{$\ket{Y}$}&$0^{N-2}1^{2}$&$\braket{A}{X}$&$\braket{A}{Y}$\\
&&$0^{N-1}2^{1}$&1&0\\
\br
\endtab

%If $\{\ket{A},\ket{B}\}$ is chosen as the working basis, i.e.,
%\begin{align*}
%\ket{3,0^{N-2}1^{1}2^{1},[N-1,1];2,1}=&\ket{A},\\
%\ket{3,0^{N-2}1^{1}2^{1},[N-1,1];2,2}=&\ket{B},
%\end{align*}
%then the coefficients in the coupling between states with $\tilde\mu\in\{0^{N-2}1^{2},0^{N-1}2^{1}\}$ and $\tilde\mu'=0^{N-2}1^{1}2^{1}$ are
%\begin{equation}
%\begin{aligned}
%C_{0^{N-2}1^{1}2^{1},0^{N-2}1^{2}}^{[N-1,1]}(1,1)=&1,\\
%C_{0^{N-2}1^{1}2^{1},0^{N-2}1^{2}}^{[N-1,1]}(1,2)=&0,\\
%C_{0^{N-2}1^{1}2^{1},0^{N-1}2^{1}}^{[N-1,1]}(1,1)=&\braket{X}{A},\\
%C_{0^{N-2}1^{1}2^{1},0^{N-1}2^{1}}^{[N-1,1]}(1,2)=&\braket{X}{B}.
%\end{aligned}
%\end{equation}
%On the other hand, the choice
%\begin{align*}
%\ket{3,0^{N-2}1^{1}2^{1},[N-1,1];2,1}=&\ket{X},\\
%\ket{3,0^{N-2}1^{1}2^{1},[N-1,1];2,2}=&\ket{Y},
%\end{align*}
%yields the coefficients
%\begin{equation}
%\begin{aligned}
%C_{0^{N-2}1^{1}2^{1},0^{N-2}1^{2}}^{[N-1,1]}(1,1)=&\braket{A}{X},\\
%C_{0^{N-2}1^{1}2^{1},0^{N-2}1^{2}}^{[N-1,1]}(1,2)=&\braket{A}{Y},\\
%C_{0^{N-2}1^{1}2^{1},0^{N-1}2^{1}}^{[N-1,1]}(1,1)=&1,\\
%C_{0^{N-2}1^{1}2^{1},0^{N-1}2^{1}}^{[N-1,1]}(1,2)=&0.
%\end{aligned}
%\end{equation}

In summary, the full symmetrized blocks of the interaction Hamiltonian are given by
\begin{equation}
\eqalign{
\hat{H}_{\textrm{int},n_\textrm{exc}}^{(\blambda,\mathbf{y})}\\
=\sum_{\tilde\mu:\bmu\trianglelefteq\blambda}\sum_{\tilde\mu':\bmu'\trianglelefteq\blambda}\sum_{i=1}^{K_{\boldsymbol{\lambda\mu}}}\sum_{i'=1}^{K_{\boldsymbol{\lambda\mu'}}}g_{\tilde\mu'\tilde\mu}B^{(n_\textrm{exc})}_{\tilde\mu',\tilde\mu}L_{\tilde\mu',\tilde\mu}^{(\blambda)}C_{\tilde\mu',\tilde\mu}^{(\blambda)}(\mathbf{w}'_{i'},\mathbf{w}_i)\ket{n_\textrm{exc},\tilde\mu',\blambda;\mathbf{y},\mathbf{w}'_{i'}}\bra{n_\textrm{exc},\tilde\mu,\blambda;\mathbf{y},\mathbf{w}_i}.
}
\end{equation}

\section{Algorithm for Hamiltonian separation}\label{sec:algor}
In this section we lay out the specific steps to write the block diagonalized Hamiltonian of a given excitation manifold.
\begin{enumerate}
\item Define  $n_\textrm{exc}$, the manifold.
\item List all relevant $\tilde\mu$, the  spectral configurations.
\begin{enumerate}
\item Enumerate all the  partitions, $\bnu=[v_1,v_2,\ldots,v_k]$, of integers from 0 to $n_\textrm{exc}$. These must amount to $\lvert\tilde\mu_{n_\textrm{exc}}\rvert$ as calculated with  \eref{eq:specconfpart} and \eref{eq:partseries}.
\item Gather the unique elements in each partition and their corresponding multiplicities. These are the excited levels, $1\leq v\leq r$, and their populations, $n_{\tilde\mu}(v)$, respectively.
\item Write the spectral configurations according to  \eref{eq:specconf}, taking into account that
\begin{equation}n_{\tilde\mu}(0)=
\begin{cases}
N&\text{if}\enskip \bnu=[0]\\
N-k&\text{otherwise}
\end{cases}.
\end{equation}
\end{enumerate}
\item Calculate all the relevant bare energies, $\varepsilon_{n_\textrm{exc}}^{(\tilde\mu)}$, with  \eref{eq:barenerg}.
\item List all the allowed $\blambda$, the symmetry defining partitions.
\begin{enumerate}
\item Enumerate the spectral configurations as $\bmu$, i.e., as partitions in regular form.
\item Discard redundancies. The remaining partitions are the possible $\blambda$. Their number must add up to $\lvert\lambda_{n_\textrm{exc}}\rvert$ as calculated with  \eref{eq:sympart} and \eref{eq:sympartseries}.
\end{enumerate}
\item Determine the composition of each $\tilde\mu$ in terms of $\blambda$ according to Young's rule in \eref{eq:youngrule}.
\begin{enumerate}
\item Find out the dominance relations among partitions according to \eref{eq:dominance}.
\item Compute the pertinent Kostka numbers ($K_{\blambda\bmu}$).
\end{enumerate}
\item Write the diagonal of $\hat{H}_{n_\textrm{exc}}^{(\blambda)}$.
\begin{enumerate}
\item Write the diagonal of $\hat{H}_{n_\textrm{exc}}^{(\blambda,1)}$.
\begin{enumerate}
\item Identify all the $\tilde\mu$ to which $\blambda$ contributes.
\item Collect the bare energies with the $\tilde\mu$ identified above and list each of them $K_{\blambda\bmu}$ times.
\end{enumerate}
\item Calculate the dimension of the representation, $\dim(S^{\blambda})$, according to the hook-length formula  \eref{eq:hooklength}.
\item  By following
\begin{equation}
\mathbf{H}_{n_\textrm{exc}}^{(\blambda)}=\mathbf 1_{\dim(S^{\blambda})}\otimes\mathbf{H}_{n_\textrm{exc}}^{(\blambda,1)},
\end{equation}
register the blocks of the Hamiltonian.
\end{enumerate}
\item Calculate the couplings.
\begin{enumerate}
\item Among the available spectral configurations, determine all pairings $\{\tilde\mu',\tilde\mu\}$ such that \begin{equation}\left\lvert\sum_{v=1}^{n_\textrm{exc}}v\left[n_{\tilde\mu'}(v)-n_{\tilde\mu}(v)\right]\right\rvert=1.\end{equation} The number of pairings inside a given manifold must be $n_\textrm{int}(n_\textrm{exc})$ as calculated with  \eref{eq:numpairs}.
\item For each pair
\begin{enumerate}
\item Identify the the levels involved in the transition, $v^*$ and $v^*+s$, as well as the respective transition dipole moment, and calculate $g_{\tilde\mu',\tilde\mu}$ with  \eref{eq:coupconst}.
\item Compute the contribution from the transition in the EM mode, $B_{\tilde\mu',\tilde\mu}^{n_\textrm{exc}}$, with  \eref{eq:couphot}.
\end{enumerate}
\item Write the off-diagonal terms of $\hat{H}_{n_\textrm{exc}}^{(\blambda,1)}$.
For each pair $\{\tilde\mu',\tilde\mu\}$ within each block of symmetry $\blambda$:
\begin{enumerate}
\item Evaluate the contributions from the transition in the space of emitters, $L_{\tilde\mu',\tilde\mu}^{(\blambda)}$, according to  \eref{eq:angmomcoup}. This can be accomplished by brute-force computation of  \eref{eq:angmomm} and \eref{eq:angmomj}, or with the strategies described in Appendix \ref{ap:calcL}.
\item If $K_{\blambda\bmu'}>1$
\begin{enumerate}
\item Construct SABEs, $\ket{n_\textrm{exc},\tilde\mu_i,\blambda;\mathbf{y}_1,\mathbf{w}_1}$, for all $\tilde\mu_i$ connected to $\tilde\mu'$ in a convenient basis.
\item Apply $\hat{J}_{+}^{\tilde\mu',\tilde\mu_i}$ to generate a basis.
\item Orthogonalize the basis.
\item Apply $\hat{J}_{-}^{\tilde\mu',\tilde\mu_i}$ to the elements of the basis and extract the quantities $L_{\tilde\mu'\tilde\mu}^{(\blambda)}C_{\tilde\mu'\tilde\mu}^{(\blambda)}(\mathbf{w}',\mathbf{w})$ using  \eref{eq:lccoup}.
\end{enumerate}
\item Calculate the coupling matrix element according to  \eref{eq:offdiag}.
\end{enumerate}
\end{enumerate}
\item Collect the calculated couplings in their respective blocks.
\end{enumerate}

\section{Worked examples}\label{sec:examp}
\subsection{The triply excited manifold}
In this section we illustrate the implementation of the algorithm when dealing with the system when it holds three quanta, i.e., $n_\textrm{exc}=3$.

\Tref{tab:tripham} compiles the quantities computed according to steps $1-4$ of the algorithm in the previous section, as well as the dimensions of the irreps in step 6.b.
\begin{table}
\tiny
\caption{\label{tab:tripham}Relevant partitions, spectral configurations, excitations in the EM mode, bare energies, and module dimensions calculated in the block diagonalization of $\hat{H}_3(N)$. }
\begin{tabular*}{\textwidth}{@{}l@{\extracolsep{0pt plus 12pt}}r*{2}{@{\extracolsep{0pt plus 12pt}}l}@{\extracolsep{0pt plus 12pt}}c*{2}{@{\extracolsep{0pt plus 12pt}}l}@{\extracolsep{0pt plus 12pt}}r*{2}{@{\extracolsep{0pt plus 12pt}}l}}
%\begin{indented}
%\item[]\begin{tabular}{*{10}{c}}
\br
$n$&$p(n)$&$\bnu$&$\tilde\mu$&$v_0^{(3,\tilde\mu)}$&{$\varepsilon_{3}^{(\tilde\mu)}$}&$\bmu$&$q(n)$&$\enskip\blambda$&$\dim(S^{\blambda})$\\
\mr
0&1&$[0]$&$0^N$&3&$\frac{7}{2}\hbar\omega+N\varepsilon_0$&$[N]$&1&$[N]$&1\\
1&1&$[1]$&$0^{N-1}1^{1}$&2&$\frac{5}{2}\hbar\omega+(N-1)\varepsilon_0+\varepsilon_1$&$[N-1,1]$&1&$[N-1,1]$&$N-1$\\
{2}&{2}&$[2]$&$0^{N-1}2^{1}$&1&$\frac{3}{2}\hbar\omega+(N-1)\varepsilon_0+\varepsilon_2$&$[N-1,1]$&{1}&-&-\\
&&$[1,1]$&$0^{N-2}1^{2}$&1&$\frac{3}{2}\hbar\omega+(N-2)\varepsilon_0+2\varepsilon_1$&$[N-2,2]$&&$[N-2,2]$&$N(N-3)/2$\\
{3}&{3}&$[3]$&$0^{N-1}3^{1}$&0&$\frac{\hbar\omega}{2}+(N-1)\varepsilon_0+\varepsilon_3$&$[N-1,1]$&{2}&-&-\\
&&$[2,1]$&$0^{N-2}1^{1}2^{1}$&0&$\frac{\hbar\omega}{2}+(N-2)\varepsilon_0+\varepsilon_1+\varepsilon_2$&$[N-2,1,1]$&&$[N-2,1,1]$&$(N-1)(N-2)/2$\\
&&$[1,1,1]$&$0^{N-3}1^{3}$&0&$\frac{\hbar\omega}{2}+(N-3)\varepsilon_0+3\varepsilon_1$&$[N-3,3]$&&$[N-3,3]$&$N(N-1)(N-5)/6$\\
\mr
&$\left\lvert\tilde\mu_{3}\right\rvert=7$&&&&&&{$\left\lvert\lambda_{3}\right\rvert=5$}&\\
\br
\end{tabular*}
%\end{indented}
\end{table}

In \tref{tab:kostka}, we show the Kostka numbers that indicate the composition of the spaces spanned by wavefunctions with the same $\tilde\mu$ in terms of symmetrized subspaces labeled by $\blambda$. Notice that $K_{\blambda\bmu}=0$ implies that $\blambda\ntrianglerighteq\bmu$.
\Table{\label{tab:kostka}Kostka numbers, $K_{\blambda\bmu}$, relating the permutation and Specht modules that appear in the triply excited manifold.}
\br
&\multicolumn{5}{c}{$\bmu$}\\
%\backslashbox{$\blambda$}{$\bmu$}&$[N]$&$[N-1,1]$&$[N-2,2]$&$[N-2,1,1]$&$[N-3,3]$\\
$\blambda$&$[N]$&$[N-1,1]$&$[N-2,2]$&$[N-2,1,1]$&$[N-3,3]$\\
\mr
%$\blambda$&\multicolumn{5}{c}{$K_{\boldsymbol{\lambda\mu}}$}\\
$[N]$&1&1&1&1&1\\
$[N-1,1]$&0&1&1&2&1\\
$[N-2,2]$&0&0&1&1&1\\
$[N-2,1,1]$&0&0&0&1&0\\
$[N-3,3]$&0&0&0&0&1\\
\br
\endtab

The factors required to calculate the off-diagonal elements, as prescribed by step 7, up until 7.c.i, are on display in \tref{tab:couplings}. Step 7.c.ii was worked in detail in the discussion leading to \tref{tab:cproj}. What is left to this section are the explicit forms of the overlaps:
\begin{equation*}
\braket{X}{A}=\left(\frac{N-2}{2(N-1)}\right)^{1/2},
\end{equation*}
and
\begin{equation*}
\braket{X}{B}=\left(\frac{N}{2(N-1)}\right)^{1/2}=\braket{A}{Y}.
\end{equation*}
\Table{\label{tab:couplings}Energy levels involved and contributions from transitions in the EM and emitter modes to the couplings between spectral configurations in the triply excited manifold.}
\br
$\tilde\mu$&$\tilde\mu'$&$v^*$&$v^*+s$&$B_{\tilde\mu',\tilde\mu}^{(3)}$&$L_{\tilde\mu',\tilde\mu}^{[N]}$&$L_{\tilde\mu',\tilde\mu}^{[N-1,1]}$&$L_{\tilde\mu',\tilde\mu}^{[N-2,2]}$\\
\mr
$0^{N}$&$0^{N-1}1^{1}$&0&1&$\sqrt{3}$&$\sqrt{N}$&\\
$0^{N-1}1^{1}$&$0^{N-2}1^{2}$&0&1&$\sqrt{2}$&$\sqrt{2(N-1)}$&$\sqrt{N-2}$&\\
$0^{N-1}1^{1}$&$0^{N-1}2^{1}$&1&2&$\sqrt{2}$&$1$&$1$&\\
$0^{N-2}1^{2}$&$0^{N-3}1^{3}$&0&1&1&$\sqrt{3(N-2)}$&$\sqrt{2(N-3)}$&$\sqrt{N-4}$\\
$0^{N-2}1^{2}$&$0^{N-2}1^{1}2^{1}$&1&2&1&$\sqrt{2}$&$\sqrt{2}$&$\sqrt{2}$\\
$0^{N-1}2^{1}$&$0^{N-2}1^{1}2^{1}$&0&1&1&$\sqrt{N-1}$&$\sqrt{N-1}$&\\
$0^{N-1}2^{1}$&$0^{N-3}3^{1}$&2&3&1&1&1&\\
\br
\endtab

With all these elements, the Hopfield-Bogoliubov forms of the block-diagonalized Hamiltonian are
%\begin{figure*}
%\tiny
%\begin{center}
\numparts
\begin{eqnarray}
\fl\mathbf H_3^{[N]}={\begin{pmatrix}

\begin{tikzpicture}[
level/.style={very thick},
trans/.style={<->,shorten >=2pt,shorten <=2pt,>=stealth},
]\tiny
\node (N0) at (3,0) {$\varepsilon_{3}^{(0^{N})}$};
\draw[level] (2.5,.25) -- (3.5,.25);
\node (N1) at (4,-1) {$\varepsilon_{3}^{(0^{N-1}1^{1})}$};
\draw[level] (3.5,-.75) -- (4.5,-.75);
\node (N2) at (5,-2) {$\varepsilon _{3}^{(0^{N-2}1^{2})}$};
\draw[level] (4.5,-1.75) -- (5.5,-1.75);
\node (M2) at (3,-4) {$\varepsilon _{3}^{(0^{N-1}2^{1})}$};
\draw[level] (2.5,-3.75) -- (3.5,-3.75);
\node (N3) at (6,-3) {$\varepsilon _{3}^{(0^{N-3}1^{3})}$};
\draw[level] (5.5,-2.75) -- (6.5,-2.75);
\node (M12) at (4,-5) {$\varepsilon _{3}^{(0^{N-2}1^{1}2^{1})}$};
\draw[level] (3.5,-4.75) -- (4.5,-4.75);
\node (M3) at (3,-6) {$\varepsilon _{3}^{(0^{N-1}3^{1})}$};
\draw[level] (2.5,-5.75) -- (3.5,-5.75);
\draw [trans] (N0) edge node[rotate=55,above=.3em,right=-.2em]{$g_{10}\sqrt {3N}$} (N1);
\draw [trans]  (N1) edge node[rotate=55,above=.3em,right=-.2em] {$g_{10}2\sqrt {N - 1}$} (N2);
\draw [trans]  (N1) edge node[rotate=-25,right] {$g_{21}\sqrt 2$} (M2);
\draw [trans]  (N2) edge node[rotate=55,above=.3em,right=-.2em] {$g_{10}\sqrt {3\left( {N - 2} \right)}$} (N3);
\draw [trans]  (N2) edge node[rotate=-25,right] {$g_{21}\sqrt{2}$} (M12);
\draw [trans]  (M2) edge node[rotate=55,above=.3em,right=-.2em] {$g_{10}\sqrt {N - 1}$} (M12);
\draw [trans]  (M2) edge node[left] {$g_{32}$} (M3);
\end{tikzpicture}
\end{pmatrix}},\\
\fl\mathbf H_3^{[N-1,1]}=\mathbf{1}_{N-1}\otimes{\begin{pmatrix}
\begin{tikzpicture}[
level/.style={very thick},
trans/.style={<->,shorten >=2pt,shorten <=2pt,>=stealth},
]
\tiny
\node (N1) at (3,0) {$\varepsilon_{3}^{(0^{N-1}1^{1})}$};
\draw[level] (2.5,.25) -- (3.5,.25);
\node (N2) at (4,-1) {$\varepsilon _{3}^{(0^{N-2}1^{2})}$};
\draw[level] (3.5,-.75) -- (4.5,-.75);
\node (N3) at (5,-2) {$\varepsilon _{3}^{(0^{N-3}1^{3})}$};
\draw[level] (4.5,-1.75) -- (5.5,-1.75);
\node (M2) at (2,-4) {$\varepsilon _{3}^{(0^{N-1}2^{1})}$};
\draw[level] (1.5,-3.75) -- (2.5,-3.75);
\node (M12a) at (3,-4.8) {};
\draw[level] (2.5,-4.75) -- (3.5,-4.75);
\node (M12) at (4,-5) {$\varepsilon _{3}^{(0^{N-2}1^{1}2^{1})}$};
\node (M12b) at (5,-4.7) {};
\draw[level] (4.5,-4.75) -- (5.5,-4.75);
\node (M3) at (2,-6) {$\varepsilon _{3}^{(0^{N-1}3^{1})}$};
\draw[level] (1.5,-5.75) -- (2.5,-5.75);
\draw [trans]  (N1) edge node[rotate=55,above=.3em,right=-.2em] {$g_{10}\sqrt {2\left(N - 1\right)}$} (N2);
\draw [trans]  (N1) edge node[rotate=-25,right] {$g_{21}\sqrt 2$} (M2);
\draw [trans]  (N2) edge node[rotate=55,above=.3em,right=-.2em] {$g_{10}\sqrt {2\left( {N - 2} \right)}$} (N3);
\draw [trans]  (N2) edge node[rotate=-25,right] {$g_{21}\sqrt{2}$} (M12a);
\draw [trans]  (M2) edge node[rotate=55,above=.3em,right=-.2em] {$g_{10}\sqrt {\frac{N}{2}- 1}$} (M12a);
%\draw [trans]  (N2) edge node[rotate=-25,right] {$g_{21}\sqrt{2}$} (M12b);
\draw [trans]  (M2) edge node[rotate=55,above=.3em,right=-.2em] {$g_{10}\sqrt {\frac{N}{2}}$} (M12b);
\draw [trans]  (M2) edge node[left] {$g_{32}$} (M3);
\end{tikzpicture}
\end{pmatrix}},\label{eq:ham311}\\
\fl\mathbf H_3^{[N-2,2]}=\mathbf{1}_{N(N-3)/2}\otimes {\begin{pmatrix}
\begin{tikzpicture}[
level/.style={very thick},
trans/.style={<->,shorten >=2pt,shorten <=2pt,>=stealth},
]
\tiny
\node (N2) at (3,0) {$\varepsilon _{3}^{(0^{N-2}1^{2})}$};
\draw[level] (2.5,.25) -- (3.5,.25);
\node (N3) at (4,-1) {$\varepsilon _{3}^{(0^{N-3}1^{3})}$};
\draw[level] (3.5,-.75) -- (4.5,-.75);
\node (M12) at (2,-2) {$\varepsilon _{3}^{(0^{N-2}1^{1}2^{1})}$};
\draw[level] (1.5,-1.75) -- (2.5,-1.75);
\draw [trans]  (N2) edge node[rotate=55,above=.3em,right=-.2em] {$g_{10}\sqrt {2\left( {N - 2} \right)}$} (N3);
\draw [trans]  (N2) edge node[rotate=-25,right] {$g_{21}\sqrt{2}$} (M12);
\end{tikzpicture}
\end{pmatrix}},\\
\fl\mathbf H_3^{[N-3,3]}=\mathbf{1}_{N(N-1)(N-5)/6}\otimes\begin{pmatrix}\varepsilon_{3}^{(0^{N-3}1^{3})}\end{pmatrix},
\end{eqnarray}
and
\begin{eqnarray}
\fl\mathbf H_3^{[N-2,1,1]}=\mathbf{1}_{(N-1)(N-2)/2}\otimes\begin{pmatrix}\varepsilon_{3}^{(0^{N-2}1^{1}2^{1})}\end{pmatrix},
\end{eqnarray}
\endnumparts
%\end{landscape}
where $\mathbf{1}_n$ is the $n\times n$ identity matrix. The matrices above are displayed in a diagrammatic form such that the horizontal lines represent energy levels, and therefore their labels correspond to diagonal elements, while the double-headed arrows indicate the couplings with their labels corresponding to off-diagonal matrix elements. In constructing  \eref{eq:ham311}, the couplings involving the states with $\tilde\mu=0^{N-2}1^{1}2^{1}$ were worked out with the basis $\{\ket{A},\ket{B}\}$.

In the case of the totally symmetric subspace, we can recognize a ladder of four superradiantly coupled levels, corresponding to the Dicke states. These are weakly connected to a two-level superradiant interaction between the SABEs with one excitation in $v=2$. In turn, this interaction connects weakly with the remaining SABE in which the third excited state is populated. For the $N-1$-degenerate subspace with $\blambda=[N-1,1]$, the superradiantly connected Dicke states form a three-level system in $\Xi$ configuration. They weakly couple to a $\Lambda$ three-level system produced by the degeneracy introduced by the two-fold contribution of $\tilde\mu={0^{N-2}}{1^1}{2^1}$ to this symmetry. As in the case above, the remaining SABE interacts only weakly with this array. For the symmetry $\blambda=[N-2,2]$ the subspace corresponds to a three-level system with configuration $\Lambda$ and degeneracy degree of $N(N-3)/2$. However, only one of the couplings is superradiant while the other is weak. The remaining subspaces with symmetries $[N-3,3]$ and $[N-2,1,1]$ are dark, and therefore there are no interactions among the states with degeneracy degrees $N(N-1)(N-5)/6$ and $(N-1)(N-2)/2$, respectively.

\subsection{Matrices for lower manifolds}
Finally, we present the Hopfield-Bogoliubov form of the Hamiltonian operators for $0\leq n_\textrm{exc}\leq2$. Their properties have been exhaustively discussed elsewhere.
\begin{equation}
\mathbf H_0^{[N]}=\begin{pmatrix}\varepsilon_{0}^{(0^{N})}\end{pmatrix},
\end{equation}
\numparts\label{eq:1xman}\begin{eqnarray}
\mathbf{H}_1^{[N]}=
\begin{pmatrix}
\varepsilon_{1}^{(0^{N})}&g_{10}\sqrt{N}
\\ g_{01}\sqrt{N}&\varepsilon_{1}^{(0^{N-1}1^{1})}
\end{pmatrix},\\
\mathbf{H}_1^{[N-1,1]}=\mathbf{1}_{N-1}\otimes\begin{pmatrix}\varepsilon_{1}^{(0^{N-1}1^{1})}\end{pmatrix},
\end{eqnarray}\endnumparts

\numparts
\begin{eqnarray}
\mathbf H_2^{[N]}=
\begin{pmatrix}
\varepsilon_{2}^{(0^{N})}&g_{10}\sqrt{2N}&0&0\\
g_{01}\sqrt{2N}&\varepsilon_{2}^{(0^{N-1}1^{1})}&g_{10}\sqrt{2(N-1)}&g_{21}\\
0&g_{01}\sqrt{2(N-1)}&\varepsilon_{2}^{(0^{N-2}1^{2})}&0\\
0&g_{12}&0&\varepsilon_{2}^{(0^{N-1}2^{1})}\end{pmatrix},\\
\mathbf H_2^{[N-1,1]}=\mathbf{1}_{N-1}\otimes \begin{pmatrix}
\varepsilon_{2}^{(0^{N-1}1^{1})}&g_{10}\sqrt{N-2}&g_{21}\\
g_{01}\sqrt{N-2}&\varepsilon_{2}^{(0^{N-2}1^{2})}&0\\
g_{12}&0&\varepsilon_{2}^{(0^{N-1}2^{1})}
\end{pmatrix}\\
\mathbf H_2^{[N-2,2]}=\mathbf{1}_{N(N-3)/2}\otimes\begin{pmatrix}\varepsilon_{2}^{0^{N-2}1^{2}}\end{pmatrix}.
\end{eqnarray}
\endnumparts

\section{Properties of eigenstates.}\label{sec:obser}

To showcase the usefulness of the formalism, in this section, we analyze the behavior of the energy spectrum and photon content of the eigenstates as a function of parameters in the Hamiltonian, such as anharmonicity, intensity of coupling and detuning.

If each of the emitters is a harmonic oscillator (HO), we can define
\begin{equation}
\hat{a}_i=\hat\sigma_{v-1,v}^{(i)}\qquad v>0,
\end{equation}
and the eigenstates can be represented as excitations of the polariton modes:
\begin{equation}\label{eq:harmeigst}
\ket{v(\textrm{UP})v(\textrm{LP})v(\textrm{D}_1)\ldots v(\textrm{D}_{N-1})}=(a_{+}^\dagger)^{v(\textrm{UP})}(a_{-}^\dagger)^{v(\textrm{LP})}\prod_{k=1}^{N-1}(\hat{a}_{\textrm{D}(k)})^{v(\textrm{D}_k)}\ket{0},
\end{equation}
where
\numparts
\begin{eqnarray}
\hat{a}_{\pm}^\dagger=\frac{1}{\sqrt{2\Omega_{10}}}\left(\pm\sqrt{\Omega_{10}\pm\Delta}\hat{a}_0^\dagger+\sqrt{\frac{\Omega_{10}\mp\Delta}{N}}\sum_{i=1}^N\hat{a}_i^\dagger\right),\\
\hat{a}_{\textrm{D}(k)}=\frac{1}{\sqrt{k}}\left(\sqrt{k-1}\hat{a}_k^\dagger-\frac{1}{\sqrt{k-1}}\sum_{i=1}^{k-1}\hat{a}_i^\dagger\right).
\end{eqnarray}
\endnumparts
The quantities $\Delta=\omega-(\varepsilon_1-\varepsilon_0)/\hbar$, and $\Omega_{10}=[\Delta^2+4(g_{10}/\hbar)^2N]^{1/2}$ are the detuning and Rabi frequency, respectively, and the labels UP, LP and D stand for \emph{upper}, \emph{lower} and \emph{dark polaritons}, respectively.

When anharmonicity is considered, the excitations in the polaritonic modes are no longer good quantum numbers; however, the actual eigenstates are similar enough to those in the harmonic case that the labels introduced in \eref{eq:harmeigst} can still be consistent with some of the features displayed by these states. We introduce anharmonicities by considering each emitter as a Morse oscillator (MO) \cite{Morse1929}, i.e., the single-emitter Hamiltonians include a potential energy function of the form
\begin{equation}
V(R)=D_e\left(1-\textrm{e}^{-a(R-R_e)}\right)^2,
\end{equation}
where $R$ is the mass-scaled length of the oscillator with value at equilibrium $R_e$, $D_e=V(\infty)-V(R_e)$ is the dissociation energy, and
\begin{equation}\label{eq:deffina}
a^2=\frac{1}{2D_e}\left.\frac{d^2 V(R)}{dR^2}\right\rvert_{R_e}.
\end{equation}
The corresponding eigenenergies are given by
\begin{equation}\label{eq:morse}
\varepsilon_v=\hbar a\left(v+\frac{1}{2}\right)\left[\sqrt{2 D_e}-\frac{\hbar a}{2}\left(v+\frac{1}{2}\right)\right]\qquad\varepsilon_v\leq D_e,
\end{equation}
with the implication that the number of bound-states is $\lfloor r_\textrm{MO}\rfloor$, where
\begin{equation}
r_\textrm{MO}=\frac{\sqrt{2D_e}}{\hbar a}-\frac{1}{2};
\end{equation}
therefore, the potential becomes harmonic as $D_e\to\infty$ ($a\to0$) \cite{DahlSpringborg1988}.

From \eref{eq:deffina}, it follows that, in the neighborhood of the equilibrium length, the emitters oscillate with frequency
\begin{equation}
\omega_\textrm{MO}=a\sqrt{2D_e}.
\end{equation}
On the other hand, the difference in energies of contiguous levels fulfill
\begin{equation}
\varepsilon_{v+1}-\varepsilon_v=\hbar \omega_\textrm{MO}-\hbar^2 a^2\left(v+1\right);
\end{equation}
therefore, this model introduces a mechanical anharmonicity characterized by $-a^2$. There is also an electric anharmonicity, which stems from the fact that \cite{LimaHornos2005}
\begin{equation}
\frac{g_{v+1,v}^\textrm{MO}}{g_{v+1,v}^\textrm{HO}}=\frac{2}{2(r_\textrm{MO}-v)-1}\left[2\frac{\sqrt{2D_e}}{\hbar a}\frac{(r_\textrm{MO}-v)(r_\textrm{MO}-v-1)}{(2r_\textrm{MO}-v)}\right]^{1/2}.
\end{equation}
It is pertinent to remark that the dipole moment operator induces transitions between all the levels in a MO; however, the model in this work focuses only on the $v\to v\pm1$ processes. 

\Fref{fig:specs} compares the eigenenergies  as functions of the Rabi splitting, $\hbar\Omega_{10}=2g_{10}\sqrt{N}$, of systems with different anharmonicities, and in which the frequency of the EM mode is resonant with the transition $0\to1$ of the emitters.  As expected, the slope of the energy as a function of the coupling intensity is proportional to the number of quanta in the non-dark modes, with the sign of the contribution being positive for UP, and negative for LP. The introduction of anharmonicity lifts the degeneracy of states with multiple excitations in the dark modes, which has some remarkable consequences.

In the harmonic case, the quantum numbers facilitate the identification of the symmetry of a state just by inspection of its multiplicity. For instance, the state $\ket{3(\textrm{UP})}$ is unique in regards of its spectral configuration; therefore, it belongs to $\blambda=[N]$. On the other hand, states of the form $\ket{1(\textrm{UP})1(\textrm{LP})1(\textrm{D}_k)}$ have multiplicity of $N-1$ and symmetry $\blambda=[N-1,1]$. Also, from the $(N-1)(N-2)/2$ states of the form $\ket{1(\textrm{LP})1(\textrm{D}_k)1(\textrm{D}_{k'})}$, one has symmetry $\blambda=[N]$ while the remaining $N(N-3)/2$ are of the $\blambda=[N-2,2]$ kind. In contrast, when anharmonicities are involved, the eigenstates can be labeled according to harmonic states of reference with the appropriate energetics. As a consequence, a state with labels $\ket{1(\textrm{LP})2(\textrm{D}_k)}$, which accounts for BEs with $\tilde{\mu}={0^{N-1}}{2^1}$, now has symmetry $\blambda=[N]$ despite belonging to a multiplet of size $N-1$. The lift in degeneracy, together with the diversity in slopes with which the eigenenergies depend on the coupling strength generate situations in which the ordering of levels changes for different coupling intensities. This might result in interesting spectroscopic observations for samples with different concentrations of emitters. Furthermore, conditions of near degeneracy modulated by light-matter coupling can result in enhancement of multiphoton absorption, as suggested in \cite{RibeiroCampos-Gonzalez-AnguloGiebinkEtAl2020} for the doubly excited manifold.
\begin{figure}[b]
\centering
\subfloat[$\blambda={[N]}$]{
\includegraphics[width=\textwidth]{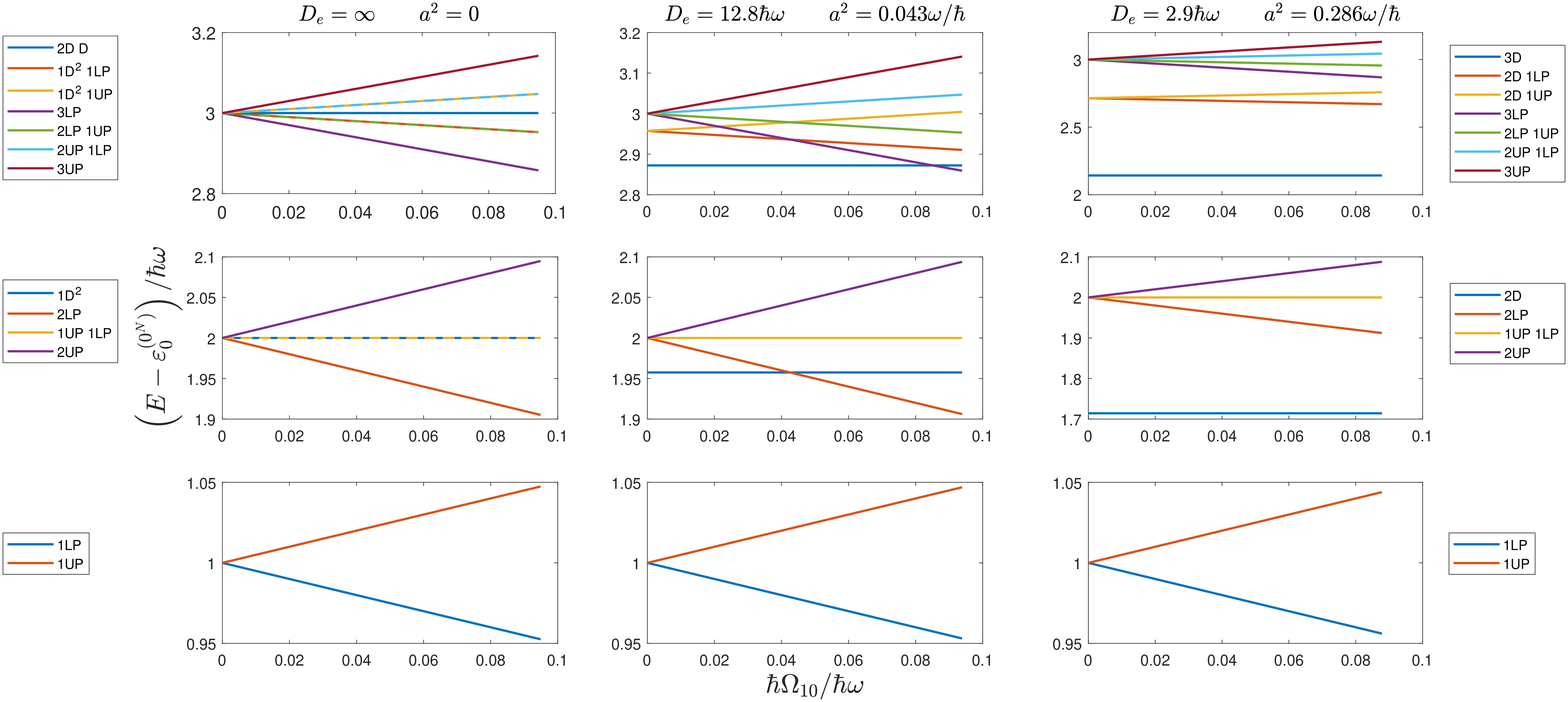}
}

\subfloat[$\blambda={[N-1,1]}$]{
\includegraphics[width=\textwidth]{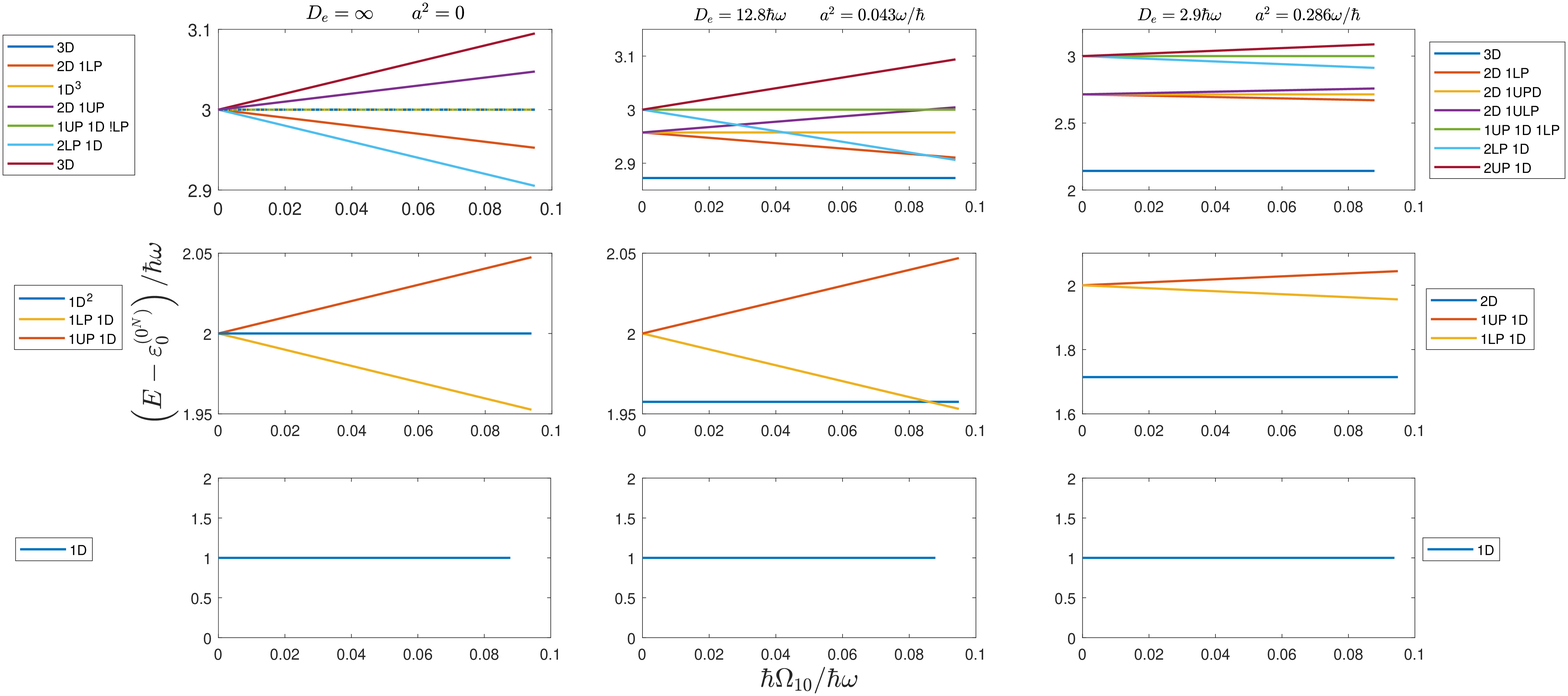}
}
\phantomcaption
\end{figure}

\clearpage

\begin{figure}\ContinuedFloat
\subfloat[$\blambda={[N-2,2]}$]{
\includegraphics[width=\textwidth]{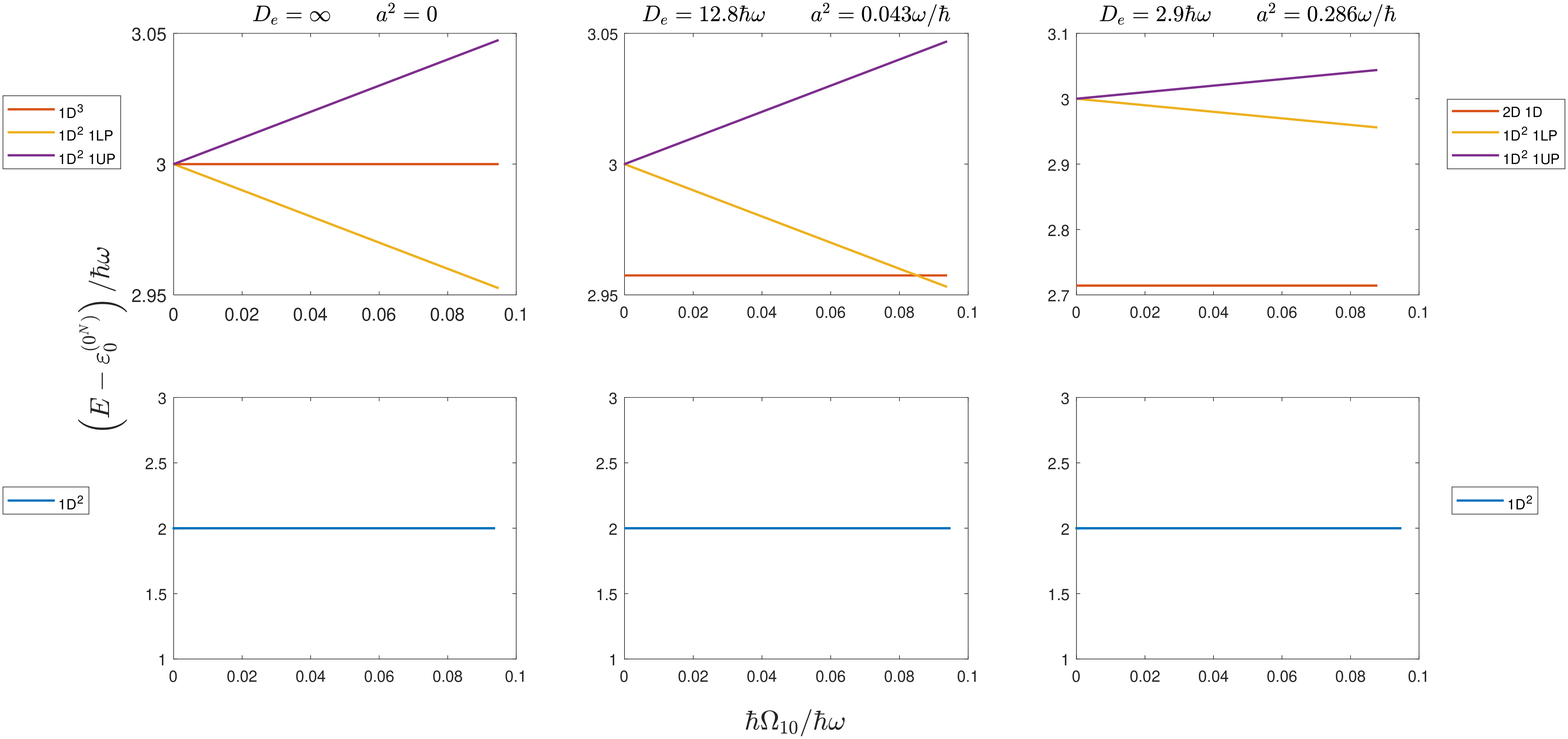}
}
\caption{Energy spectra of emitters with various anharmonicities coupled to a harmonic EM mode as a function of the coupling intensity, $\hbar\Omega_{10}$, in the weak-to-strong regime. The parameters of the Morse potential are such that the number of bound states, $r_\textrm{MO}$, are $\infty$, 24 and 4, respectively, for each of the displayed columns. The legend to the left shows the labels of states in the harmonic case, while the legend to the right labels states in the anharmonic regime. \label{fig:specs}}
\end{figure}

\begin{figure}
\includegraphics[width=\textwidth]{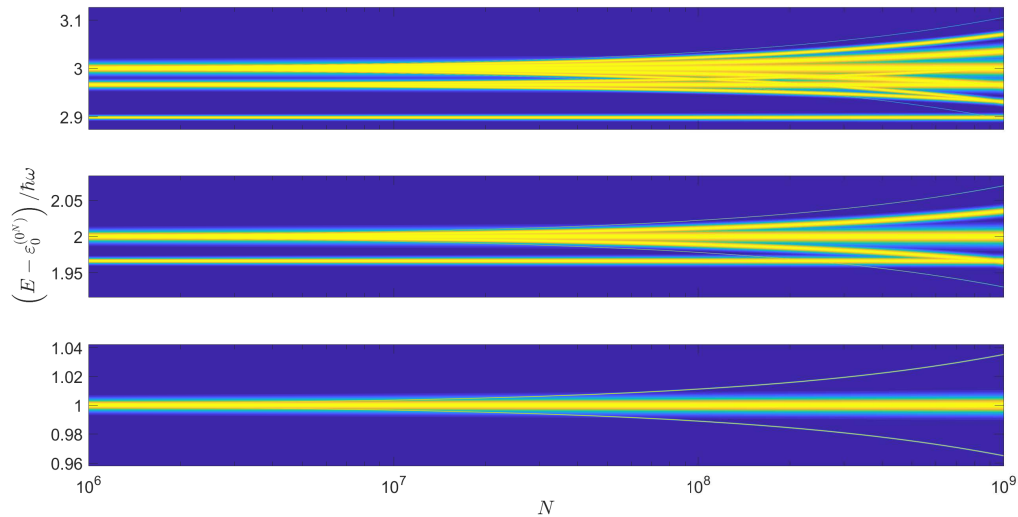}
\caption{Energy spectra of anharmonic emitters coupled to a harmonic EM mode with illustrated degeneracies as a function of the number of emitters, $N$. The Morse potential parameters are $D_e=15.25\hbar\omega$ and $a^2=0.034\omega/\hbar$. The thickness of the lines are logarithmically proportional to the degeneracies of the irreps, $\dim(S^{\blambda})$. \label{fig:degens}}
\end{figure}

In \fref{fig:degens}, the energy of the eigenstates is plotted against the number of coupled molecules with the thickness of the lines illustrating the degeneracy of each energy level.  In each case, the thickness corresponds to $\log[\dim(S^{\blambda})]$ and should not be mistaken by any broadening mechanism such as dissipation or disorder. It can be seen that both, degeneracy and energy separation, increase with the number of molecules. Furthermore, the most degenerate levels are those with constant energy. Among such states there are dark, with no component whatsoever from the EM mode, and states where $v(\textrm{UP})=v(\textrm{LP})$ in which the contributions from both polaritonic modes balance each other energetically. This observation informs that light-matter coupling might not impact processes that engage the matter component, such as chemical reactions, even when anharmonicities are taken into account. Moreover, the resolution in energies also suggests that naturally occurring broadenings shall smear bands of nearby energy levels making them effectively indistinguishable.

\begin{figure}
\includegraphics[width=\textwidth]{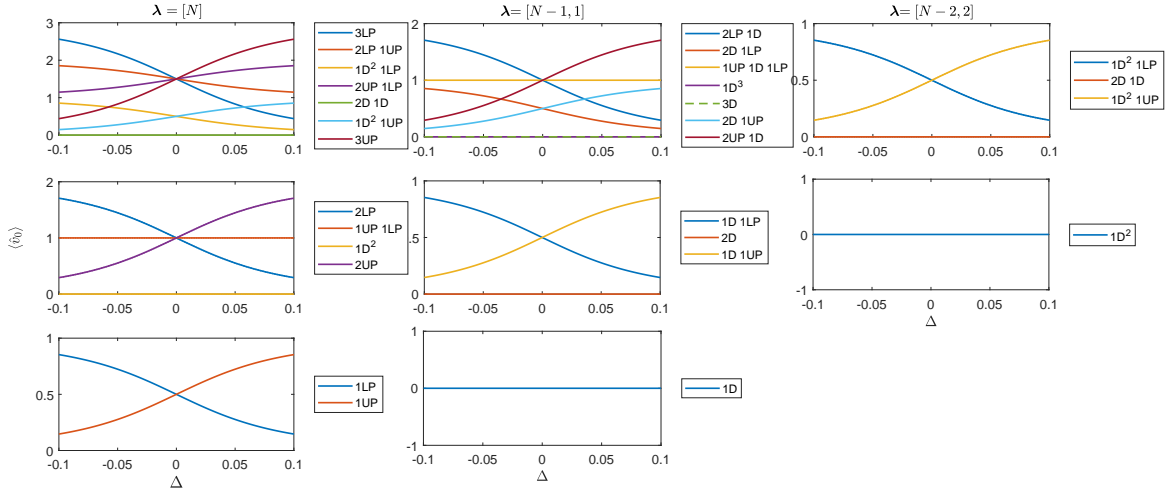}
\caption{Photon content of harmonic eigenstates as a function of detuning.\label{fig:photocon}}
\end{figure}

The formalism presented in this work also allows to calculate observables associated with operators for which the permutational symmetry holds. One clear example is the photon number operator, $\hat v_0=\hat a_0^\dagger\hat a_0$, which measures the photon contents of a given state. In the symmetrized basis, this operator is diagonal and has the form
\begin{equation}
\hat v_0=\sum_{\tilde\mu:\bmu\trianglelefteq\blambda}\sum_{i=1}^{\dim(S^{\blambda})}\sum_{j=1}^{K_{\blambda\bmu}}v_0^{(n_\textrm{exc},\tilde\mu)}\ket{n_\textrm{exc},\tilde\mu,\blambda;\mathbf{y}_i,\mathbf{w}_j}\bra{n_\textrm{exc},\tilde\mu,\blambda;\mathbf{y}_i,\mathbf{w}_j}.
\end{equation}
In \fref{fig:photocon} the photon content of the eigenstates is plotted as a function of the detuning for a fixed value of the collective coupling. The inclusion of anharmonicity should be inconsequential except for the change in the labeling of eigenstates.

\section{Conclusions.}\label{sec:conclu}

In this paper, we address the problem of the coupling between $N$ identical dipoles, each with an arbitrary spectral structure, and a harmonic electromagnetic mode confined in a cavity. We have introduced tools from Group Theory that capitalize from the permutational symmetry of the system to simplify the time-independent Schrödinger equation. In the symmetry-adapted basis, the Hamiltonian breaks down into manageable matrices whose dimension is independent of the number of emitters and grows subexponentially with the number of excitations. While the total number of these matrices does depend polynomially on the number of emitters, they encode highly degenerate subspaces; therefore, the effective number of matrices to diagonalize is actually small. Since the method here presented does not rely on the explicit form of the basis transformation, the off-diagonal matrix elements are constructed by taking advantage of the fact that the structure of the Hilbert space also displays the symmetry of the special unitary group, and therefore can be described with the tools from angular momentum theory. This procedure exhibits the (super)radiant character of the transitions.
We have also explored the immediate implications of including anharmonicities on the energetic and combinatorial characterization of the eigenstates. Finally, we have calculated the photon content of the eigenstates to exemplify the utility of the method in the extraction of observables associated with symmetry-preserving operators.

\ack
J.A.C.G.A  thanks Kai Schwennicke and Stephan van den Wildenberg for their useful comments and insights. J.A.C.G.A., R.F.R., and J.Y.Z. acknowledge funding support from the Defense Advanced Research Projects Agency under Award No. D19AC00011, and the Air Force Office of Scientific Research award FA9550-18-1-0289.
J.A.C.G.A. acknowleges initial support of the UC-MEXUS-CONACYT graduate scholarship ref. 235273/472318. J.Y.Z. was funded with the NSF EAGER Award CHE 1836599. 

\appendix
\section{Generating operators of the Schur-Weyl basis.\label{sec:swbasoper}}
Tables \ref{tab:swi} and \ref{tab:swii} show the excitation operators, $\hat{O}\{\tilde\mu,{\blambda};\mathbf{y},\mathbf{w}\}$ that produce the states with the same labels when applied to the global groundstate of the emitters, i.e.
\begin{equation}
\hat{O}\{\tilde\mu,{\blambda};\mathbf{y},\mathbf{w}\}\ket{0}=\mathcal{N}^{-2}\ket{\tilde\mu,{\blambda};\mathbf{y},\mathbf{w}}.
\end{equation}
The nested structure of the operators responds to the fact that the higher states are obtained through either collective excitations, or Gram-Schmidt orthogonalization of the lower ones.
\begin{landscape}
%\fulltable{\label{tab:swi}SABEs in terms of PBEs for spectral configurations with at most three excitations in the emitters.}
\begin{table}
\tiny
\caption{\label{tab:swi}SABEs in terms of PBEs for spectral configurations with at most three excitations in the emitters.}
\begin{tabular*}{\linewidth}{@{}l*{5}{l}@{\extracolsep{0pt plus 12pt}}l}
\br
$\tilde\mu$&$\blambda$&$\mathbf{y}$&$\mathbf{w}$&$\hat{O}\{\tilde\mu,\blambda;\mathbf{y},\mathbf{w}\}$&$\mathcal{N}^{-2}$&Notes\\
\mr

$0^{N}$&$[N]$&0&0&$1$&1&\\
\ms
$0^{N-1}v^{1}$&$[N]$&0&0&$\displaystyle\sum\limits_{i=1}^N\hat{\sigma}_{v,0}^{(i)}$&$N$&$v\in\{1,2,3\}$\\
\ms
&$[N-1,1]$&$k$&${v^1}$&$(k-1)\hat{\sigma}_{v,0}^{(k)}-\hat{O}\{0^{k-2}1^{1},[k-1];0,0\}$&$k(k-1)$&$2\leq k\leq N$\\
\ms
$0^{N-2}1^{2}$&$[N]$&0&0&$\displaystyle2\sum\limits_{i=1}^{N-1}\sum\limits_{j=i+1}^{N}\hat{\sigma}_{1,0}^{(i)}\hat{\sigma}_{1,0}^{(j)}$&$2N(N-1)$\\
\ms
&$[N-1,1]$&$k$&${1^1}$&$\begin{aligned}
(k-2)&\hat{\sigma}_{1,0}^{(k)}\hat{O}\{0^{k-2}1^{1},[k-1];0,0\}-\hat{O}\{0^{k-3}1^{2},[k-1];0,0\}\\
+&\sum\limits_{j=k+1}^{N}\hat{\sigma}_{1,0}^{(j)}\hat{O}\{0^{k-1}1^{1},[k-1,1];k,{1^1}\}%\left[(k-1)\ket{1_k 1_j}-\sum\limits_{i=1}^{k-1}\ket{1_i 1_j}\right]
\end{aligned}$&$k(k-1)(N-2)$&$2\leq k\leq N$\\
\ms
&$[N-2,2]$&$(k,\ell)$&${1^2}$&$(\ell-3)\hat{\sigma}_{1,0}^{(\ell)}\hat{O}\{0^{k-1}1^{1},[k-1,1];k,{1^1}\}-\hat{O}\{0^{\ell-3}1^{2},[\ell-2,1];k,{1^1}\}$&$k(k-1)(\ell-2)(\ell-3)$&$\begin{array}{c}
4\leq\ell\leq N\\2\leq k<\ell
\end{array}$\\
\ms
$0^{N-3}1^{3}$&$[N]$&0&0&$\displaystyle 6\sum\limits_{h=1}^{N-2}\sum\limits_{i=h+1}^{N-1}\sum\limits_{j=i+1}^N\hat{\sigma}_{1,0}^{(h)}\hat{\sigma}_{1,0}^{(i)}\hat{\sigma}_{1,0}^{(j)}$&$6N(N-1)(N-2)$&\\
\ms
&$[N-1,1]$&$k$&${1^1}$&$\begin{aligned}
(k-3)&\hat{\sigma}_{1,0}^{(k)}\hat{O}\{0^{k-3}1^{2},[k-1];0,0\}-\hat{O}\{0^{k-4}1^{3},[k-1];0,0\}\\
+&2\sum\limits_{h=k+1}^N\hat{\sigma}_{1,0}^{(h)}\left[(k-2)\hat{\sigma}_{1,0}^{(k)}\hat{O}\{0^{k-2}1^{1},[k-1];0,0\}-\hat{O}\{0^{k-3}1^{2},[k-1];0,0\}\right]\\
&\quad+2\sum\limits_{j=k+1}^{N-1}\sum\limits_{h=j+1}^{N}\hat{\sigma}_{1,0}^{(j)}\hat{\sigma}_{1,0}^{(h)}\hat{O}\{0^{k-1}1^{1},[k-1,1];k,{1^1}\}
\end{aligned}$&$2k(k-1)N(N-1)$&$2\leq k\leq N$\\
\ms
&$[N-2,2]$&$(k,\ell)$&${1^2}$&$\begin{aligned}
(\ell-4)&\hat{\sigma}_{1,0}^{(\ell)}\hat{O}\{0^{\ell-3}1^{2},[\ell-2,1];k,1\}+\sum\limits_{h=\ell+1}^{N}\hat{\sigma}_{1,0}^{(h)}\hat{O}\{0^{\ell-3}1^{2},[\ell-3,2];(k,\ell),{1^2}\}\\
-&\hat{O}\{0^{\ell-4}1^{3},[\ell-2,1];k,{1^1}\}
\end{aligned}$&$k(k-1)(\ell-2)(\ell-3)(N-4)$&$\begin{array}{c}
4\leq\ell\leq N\\2\leq k<\ell
\end{array}$\\
\ms
&$[N-3,3]$&$(k,\ell,m)$&${1^3}$&$(m-5)\hat{\sigma}_{1,0}^{(m)}\hat{O}\{0^{\ell-3}1^{2},[\ell-3,2];(k,\ell),{1^2}\}+\hat{O}\{0^{m-4}1^{3},[m-3,2];(k,\ell),{1^2}\}$&$k(k-1)(\ell-2)(\ell-3)(m-4)(m-5)$&$\begin{array}{c}
6\leq m\leq N\\4\leq\ell < m\\2\leq k<\ell
\end{array}$\\
\br
\endfulltable
%\end{tabular}
%\end{indented}
%\end{table}
%\end{turnpage}

%\begin{turnpage}
\begin{table}
\tiny
\caption{\label{tab:swii}SABEs in terms of PBEs with $\tilde\mu={0^{N-2}}{1^1}{2^1}$.}
\begin{tabular*}{\linewidth}{@{}l*{6}{l}}
%\fulltable{\label{tab:swii}SABEs in terms of PBEs with $\tilde\mu={0^{N-2}}{1^1}{2^1}$.}
\br
$\tilde\mu$&$\blambda$&$\mathbf{y}$&$\mathbf{w}$&$\hat{O}\{\tilde\mu,\blambda;\mathbf{y},\mathbf{w}\}$&$\mathcal{N}^{-2}$&Notes\\
\mr
${0^{N-2}}{1^1}{2^1}$&$[N]$&0&0&$\displaystyle 2\sum\limits_{i=1}^{N-1}\sum\limits_{j=i+1}^{N}\left(\hat{\sigma}_{2,0}^{(i)}\hat{\sigma}_{1,0}^{(j)}+\hat{\sigma}_{2,0}^{(j)}\hat{\sigma}_{1,0}^{(i)}\right)$&$4N(N-1)$&\\
\ms
&$[N-1,1]$&$k$&${1^1}$&$\begin{aligned}
(k-2)&\left(\hat{\sigma}_{2,0}^{(k)}\hat{O}\{{0^{k-2}}{1^1},[k-1];0,0\}+\hat{\sigma}_{1,0}^{(k)}\hat{O}\{{0^{k-2}}{2^1},[k-1];0,0\}\right)\\
-&\sum\limits_{j=k+1}^{N}\left(\hat{\sigma}_{1,0}^{(j)}\hat{O}\{{0^{k-2}}{2^1},[k-1];0,0\}+\hat{\sigma}_{2,0}^{(j)}\hat{O}\{{0^{k-2}}{1^1},[k-1];0,0\}\right)\\
&\quad+(k-1)\sum\limits_{j=k+1}^{N}\left(\hat{\sigma}_{2,0}^{(k)}\hat{\sigma}_{1,0}^{(j)}+\hat{\sigma}_{1,0}^{(k)}\hat{\sigma}_{2,0}^{(j)}\right)-\hat{O}\{{0^{k-3}}{1^1}{2^1},[k-1],0,0\}
\end{aligned}$&$2k(k-1)(N-2)$&$2\leq k\leq N$\\
\ms
&&&${2^1}$&$\begin{aligned}
k&\left(\hat{\sigma}_{2,0}^{(k)}\hat{O}\{{0^{k-2}}{1^1},[k-1];0,0\}-\hat{\sigma}_{1,0}^{(k)}\hat{O}\{{0^{k-2}}{2^1},[k-1];0,0\}\right)\\
&\enskip+\sum\limits_{j=k+1}^{N}\left(\hat{\sigma}_{2,0}^{(j)}\hat{O}\{{0^{k-2}}{1^1},[k-1];0,0\}-\hat{\sigma}_{1,0}^{(j)}\hat{O}\{{0^{k-2}}{2^1},[k-1];0,0\}\right)\\
&\quad+(k-1)\sum\limits_{j=k+1}^{N}\left(\hat{\sigma}_{2,0}^{(k)}\hat{\sigma}_{1,0}^{(j)}+\hat{\sigma}_{1,0}^{(k)}\hat{\sigma}_{2,0}^{(j)}\right)
\end{aligned}$&$2k(k-1)N$&$2\leq k\leq N$\\
\ms
&$[N-2,2]$&$(k,\ell)$&${1^1}{2^1}$&$\begin{aligned}
(\ell-3)&\left(\hat{\sigma}_{1,0}^{(\ell)}\hat{O}\{{0^{k-1}}{2^1},[k-1,1];k,{2^1}\}+\hat{\sigma}_{2,0}^{(\ell)}\hat{O}\{{0^{k-1}}{1^1},[k-1,1];k,{1^1}\}\right)\\
-&\hat{O}\{{0^{\ell-3}}{1^1}{2^1},[\ell-2,1];k,{1^1}\}
\end{aligned}$&$2k(k-1)(\ell-2)(\ell-3)$&$\begin{array}{c}
4\leq\ell\leq N\\2\leq k<\ell
\end{array}$\\
\ms
&$[N-2,1,1]$&$(k;\ell)$&$({1^1},{2^1})$&$\begin{aligned}
(\ell-2)&\left(\hat{\sigma}_{2,0}^{(k)}\hat{\sigma}_{1,0}^{(\ell)}+\hat{\sigma}_{1,0}^{(k)}\hat{\sigma}_{2,0}^{(\ell)}\right)+\sum\limits_{i=k+1}^{\ell-1}\left[\left(\hat{\sigma}_{2,0}^{(k)}-\hat{\sigma}_{2,0}^{(\ell)}\right)\hat{\sigma}_{1,0}^{(i)}+\left(\hat{\sigma}_{1,0}^{(\ell)}-\hat{\sigma}_{1,0}^{(k)}\right)\hat{\sigma}_{2,0}^{(i)}\right]\\
+&\left(\hat{\sigma}_{2,0}^{(k)}-\hat{\sigma}_{2,0}^{(\ell)}\right)\hat{O}\{{0^{k-1}}{1^1},[k-1];0,0\}+\left(\hat{\sigma}_{1,0}^{(\ell)}-\hat{\sigma}_{1,0}^{(k)}\right)\hat{O}\{{0^{k-1}}{2^1},[k-1];0,0\}
\end{aligned}$&$2\ell(\ell-1)$&$\begin{array}{c}
3\leq\ell\leq N\\2\leq k<\ell
\end{array}$\\
\br
\endfulltable
\end{landscape}

\section{Efficient calculation of $L_{\tilde\mu',\tilde\mu}^{(\blambda)}$\label{ap:calcL}}
In this section, we present two approaches, one analytical and other computational, to the calculation of the contribution of the collective transitions in the Hilbert space of the emitters to the couplings between states in  \eref{eq:offdiag}.

First, we introduce a by-hand method to compute individual instances of  \eref{eq:angmomcoup}. Taking advantage of the fact that $n_{\tilde\mu}(v^*+s)+n_{\tilde\mu}(v^*)=n_{\tilde\mu'}(v^*+s)+n_{\tilde\mu'}(v^*)$, we can write
\begin{equation}
L_{\tilde\mu'\tilde\mu}^{(\blambda)}=\left\{\left[n_{\tilde\mu}(v^*)-\rho_{J}\right]\left[n_{\tilde\mu'}(v^*+s)-\rho_{J}\right]\right\}^{1/2},
\end{equation}
where $\rho_J=[n_{\tilde\mu}(v^*+s)+n_{\tilde\mu}(v^*)]/2-J_{\tilde\mu'\tilde\mu}^{(\blambda)}$. When $\rho_J=0$, which is a typical situation among low excitation manifolds, the term inside the radical has an intuitive interpretation since it can be read as the product of the number of emitters in $\tilde\mu$ available for excitation with the number of emitters in $\tilde\mu'$ available for deexcitation. Furthermore, $\rho_J$ increases as $\blambda$ becomes less dominant, i.e., strays away from $[N]$; this provides with an automatic way to determine the symmetry labels for which a given coupling is worth to calculate. \tref{tab:Lbyhand} illustrates this calculation for the couplings between $\tilde\mu=0^{N-2}1^{2}$ and $\tilde\mu'=0^{N-3}1^{3}$.

\Table{\label{tab:Lbyhand}Contribution of transitions in emitter space to the coupling coefficients between  $\tilde\mu=0^{N-2}1^{2}$ and $\tilde\mu'=0^{N-3}1^{3}$. Notice that $n_{\tilde\mu}(0)=N-2$ and $n_{\tilde\mu'}(1)=3$.}
\br
$\blambda$&$J_{0^{N-3}1^{3},0^{N-2}1^{2}}^{(\blambda)}$&$\rho_J$&$L_{0^{N-3}1^{3},0^{N-2}1^{2}}^{(\blambda)}$\\
\mr
$[N]$&$\frac{N}{2}$&0&$\sqrt{3(N-2)}$\\
$[N-1,1]$&$\frac{N}{2}-1$&1&$\sqrt{2(N-3)}$\\
$[N-2,2]$&$\frac{N}{2}-2$&2&$\sqrt{N-4}$\\
$[N-3,3]$&$\frac{N}{2}-3$&3&0\\
\br
\endtab

Second, we discuss a computational method to get not only multiple values of the sought coefficients, but also the SABEs connected by a particular raising operator. Given that
\begin{equation}
\hat{J}_{+}^{(\tilde\mu',\tilde\mu)}\ket{\tilde\mu,\blambda;\mathbf{y},\mathbf{w}}=L_{\tilde\mu',\tilde\mu}^{(\blambda)}C_{\tilde\mu',\tilde\mu}^{(\blambda)}(\mathbf{w}',\mathbf{w})\ket{\tilde\mu',\blambda;\mathbf{y},\mathbf{w}'},
\end{equation}
it is possible to write

\begin{equation}\label{eq:svd}
\hat{J}_{+}^{(\tilde\mu',\tilde\mu)}=\sum_{\blambda\trianglerighteq\bmu'}\sum_{i=1}^{\dim(S^{\blambda})}\sum_{j'=1}^{K_{\boldsymbol{\lambda\mu}'}}\sum_{j=1}^{K_{\boldsymbol{\lambda\mu}}}L_{\tilde\mu',\tilde\mu}^{(\blambda)}C_{\tilde\mu',\tilde\mu}^{(\blambda)}(\mathbf{w}',\mathbf{w})\ket{\tilde\mu',\blambda;\mathbf{y}_i,\mathbf{w}'_{j'}}\bra{\tilde\mu,\blambda;\mathbf{y}_i,\mathbf{w}_j},
\end{equation}

and define
%\hat{V}_{\tilde\mu'}=&\sum_{\blambda\trianglerighteq\bmu'}\sum_{i=1}^{\dim(S^{\blambda})}\sum_{ t'=1}^{K_{\boldsymbol{\lambda\mu}'}}\ket{\tilde\mu',\blambda;y_i, t'}\bra{\tilde\mu',\blambda;y_i, t'},\\
\begin{equation}
\hat{U}_{\tilde\mu}=\sum_{\blambda\trianglerighteq\bmu}\sum_{i=1}^{\dim(S^{\blambda})}\sum_{j=1}^{K_{\boldsymbol{\lambda\mu}}}\ket{\tilde\mu,\blambda;\mathbf{y}_i,\mathbf{w}_j}\bra{\tilde\mu,\blambda;\mathbf{y}_i,\mathbf{w}_j}.
\end{equation}
The latter implies that, if the raising operator is written in the basis of the PBEs, the coefficients $L_{\tilde\mu',\tilde\mu}^{(\blambda)}C_{\tilde\mu',\tilde\mu}^{(\blambda)}(\mathbf{w}',\mathbf{w})$ are its singular values, while the unitary operators $\hat{U}_{\tilde\mu'}$ and $\hat{U}_{\tilde\mu}$ correspond to the matrices of singular vectors, which provide with the changes of basis between PBEs and SABEs. In short, singular value decomposition of $\hat{J}_{+}^{(\tilde\mu',\tilde\mu)}$ in an arbitrary basis yields the coupling coefficients as well as the symmetrized states in the subspaces of functions with the spectral configurations involved in the transition. The apparent contradiction in  \eref{eq:svd} that there are symmetry labels such that $\blambda\trianglerighteq\bmu'$ yet $\blambda\ntrianglerighteq\bmu$ is solved by recognizing the set of states $\{\ket{\tilde\mu',\lambda;y, t'}:\blambda\ntrianglerighteq\bmu\}$ as the null-space of the raising operator, i.e., $L_{\tilde\mu',\tilde\mu}^{(\blambda\ntrianglerighteq\bmu)}=0$. This feature is consistent with the fact that $\dim(M^{\bmu'})>\dim(M^{\bmu})$.

To conclude, we work the example of the coupling, in a system with $N=4$, between $\tilde\mu=0^{3}1^{1}$ and $\tilde\mu'=0^{2}1^{2}$. The subspaces and couplings are given by
\begin{equation}
\begin{pmatrix}
\ket{1_1}\\
\ket{1_2}\\
\ket{1_3}\\
\ket{1_4}
\end{pmatrix}
\xrightleftharpoons[\hat{J}_{-}^{(1,0)}]{\hat{J}_{+}^{(1,0)}}
\begin{pmatrix}
\ket{1_1 1_2}\\
\ket{1_1 1_3}\\
\ket{1_1 1_4}\\
\ket{1_2 1_3}\\
\ket{1_2 1_4}\\
\ket{1_3 1_4}
\end{pmatrix}
\end{equation}
In this basis, the raising operator is
\begin{equation}
\mathbf{J}_{+}^{(0^{2}1^{2},0^{3}1^{1})}=
\begin{pmatrix}
1&1&0&0\\
1&0&1&0\\
1&0&0&1\\
0&1&1&0\\
0&1&0&1\\
0&0&1&1
\end{pmatrix}.
\end{equation}
Its singular value decomposition yields
%\begin{widetext}
\begin{eqnarray}
\mathbf{U}_{0^{3} 1^{1}}=\frac{1}{2}
\begin{pmatrix}
1&-\sqrt{3}&0&0\\
1&\frac{1}{\sqrt{3}}&-\sqrt{\frac{8}{3}}&0\\
1&\frac{1}{\sqrt{3}}&\sqrt{\frac{2}{3}}&-\sqrt{2}\\
1&\frac{1}{\sqrt{3}}&\sqrt{\frac{2}{3}}&\sqrt{2}\\
\end{pmatrix},\\
%\begin{pmatrix}
%1/2&-\sqrt{3}/2&0&0\\
%1/2&1/\sqrt{12}&-\sqrt{2/3}&0\\
%1/2&1/\sqrt{12}&1/\sqrt{6}&-1/\sqrt{2}\\
%1/2&1/\sqrt{12}&1/\sqrt{6}&1/\sqrt{2}\\
%\end{pmatrix},\\
\mathbf{U}_{0^{2} 1^{2}}=\frac{1}{\sqrt{6}}
\begin{pmatrix}
1&-1&-\sqrt{2}&0&-\frac{1}{\sqrt{2}}&\sqrt{\frac{3}{2}}\\
1&-1&\frac{1}{\sqrt{2}}&-\sqrt{\frac{3}{2}}&\sqrt{2}&0\\
1&-1&\frac{1}{\sqrt{2}}&\sqrt{\frac{3}{2}}&-\frac{1}{\sqrt{2}}&-\sqrt{\frac{3}{2}}\\
1&1&-\frac{1}{\sqrt{2}}&-\sqrt{\frac{3}{2}}&-\frac{1}{\sqrt{2}}&-\sqrt{\frac{3}{2}}\\
1&1&-\frac{1}{\sqrt{2}}&\sqrt{\frac{3}{2}}&\sqrt{2}&0\\
1&1&\sqrt{2}&0&-\frac{1}{\sqrt{2}}&\sqrt{\frac{3}{2}}
\end{pmatrix},\\
%\hat{U}_{0^{2} 1^{2}}=&
%\begin{pmatrix}
%1/\sqrt{6}&-1/\sqrt{6}&-1/\sqrt{3}&0&-1/\sqrt{12}&1/2\\
%1/\sqrt{6}&-1/\sqrt{6}&1/\sqrt{12}&-1/2&1/\sqrt{3}&0\\
%1/\sqrt{6}&-1/\sqrt{6}&1/\sqrt{12}&1/2&-1/\sqrt{12}&-1/2\\
%1/\sqrt{6}&1/\sqrt{6}&-1/\sqrt{12}&-1/2&-1/\sqrt{12}&-1/2\\
%1/\sqrt{6}&1/\sqrt{6}&-1/\sqrt{12}&1/2&1/\sqrt{3}&0\\
%1/\sqrt{6}&1/\sqrt{6}&1/\sqrt{3}&0&-1/\sqrt{12}&1/2
%\end{pmatrix},\\
\mathbf{S}_{0^{3}1^{1}}^{0^{2}1^{2}}=
%\hat{U}_{0^{2}1^{2}}^\dagger \hat{J}_{+}^{(0^{2}1^{2},0^{3}1^{1})}\hat{U}_{0^{3}1^{1}}=&
\begin{pmatrix}
\sqrt{6}&0&0&0\\
0&\sqrt{2}&0&0\\
0&0&\sqrt{2}&0\\
0&0&0&\sqrt{2}\\
0&0&0&0\\
0&0&0&0
\end{pmatrix},
\end{eqnarray}
%\end{widetext}
where $\mathbf{S}_{\tilde\mu}^{\tilde\mu'}=\mathbf{U}_{\tilde\mu'}^\dagger \mathbf{J}_{+}^{(\tilde\mu',\tilde\mu)}\mathbf{U}_{\tilde\mu}$ is the matrix whose diagonal lists the singular values of $\mathbf{J}_{+}^{(\tilde\mu',\tilde\mu)}$. Notice that these values correspond with the expected $\sqrt{2(N-1)}$ and $\sqrt{N-2}$ with degeneracy degrees of 1 and $N-1$ respectively; furthermore, the null-space has dimension $N(N-3)/2$.

\section*{References}
%%\nocite{*}
\bibliographystyle{iopart-num}
\bibliography{bibmltc}% Produces the bibliography via BibTeX.
%\printbibliography

\end{document}